\documentclass[twoside,12pt]{article}
\usepackage{macros2e}
\usepackage{newcomnd}
\usepackage{epsf}
\usepackage{graphicx}
\usepackage{amssymb}
\usepackage{bildmachen}
\usepackage{times}

\begin{document}
\begin{titlepage}
\noindent
\renewcommand{\thefootnote}{\fnsymbol{footnote}}
\parbox{1.85cm}{\epsfxsize=1.85cm \epsfbox{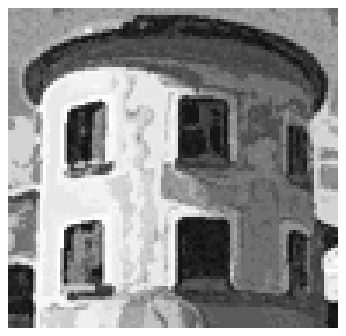}}~~ 
\parbox{1.85cm}{\epsfxsize=1.85cm \epsfbox{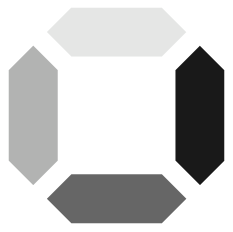}}\hfill%
\begin{minipage}{10cm}
\rightline{Uni Essen and ITP preprint}%
\rightline{cond-mat/0110011}%
\rightline{September 30, 2001}%
\end{minipage}%
\vfill 
\centerline{\sffamily\bfseries\Large Interacting Crumpled Manifolds}
\vfill
\centerline{\bf\large Henryk A.\ Pinnow$^{1}$ and Kay J\"org Wiese$^{1,2}$%
\footnote{Email: pinnow@theo-phys.uni-essen.de, wiese@itp.ucsb.edu}}
\smallskip
\centerline{\small $^{1}$ Fachbereich Physik, Universit\"at Essen,  45117 Essen,
Germany}\centerline{\small $^{2}$ ITP, Kohn Hall, University of California at Santa
Barbara, CA 93106-4030, USA}
\smallskip\smallskip

\vfill
\vspace{-5mm}
\begin{abstract}
In this article we study the effect of a $\delta $-interaction on a
polymerized membrane of arbitrary internal dimension $D$. Depending on
the dimensionality of membrane and embedding space, different physical
scenarios are observed. We emphasize on the difference of polymers
from membranes. For the latter, non-trivial contributions appear at
the 2-loop level. 
We also exploit a ``massive scheme'' inspired by
calculations in fixed dimensions for scalar field theories.
Despite the fact that these calculations are only amenable numerically,
we found that in the limit of $D\to 2$ each diagram can be evaluated 
{\em analytically}. This property extends in fact to any order in
perturbation theory, allowing for a summation of all orders. This is a
novel and 
quite surprising result. Finally,
an attempt to go beyond $D=2$ is presented. Applications to the case
of self-avoiding membranes are mentioned. 



\end{abstract}
\vspace{-5mm}
\vfill

\centerline{\em Submitted to Journal of Physics} 
\vfill

%
%
\end{titlepage}

\renewcommand{\thefootnote}{\fnsymbol{footnote}}
{
\setcounter{tocdepth}{4}
\tableofcontents} \newpage
\setcounter{page}{3}

\renewcommand{\thefootnote}{\arabic{footnote}}

\savebild{\GA}{\bildGA}{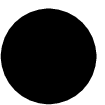}
\savebild{\GB}{\bildGB}{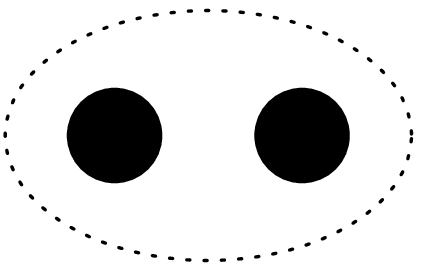}
\savebild{\GC}{\bildGC}{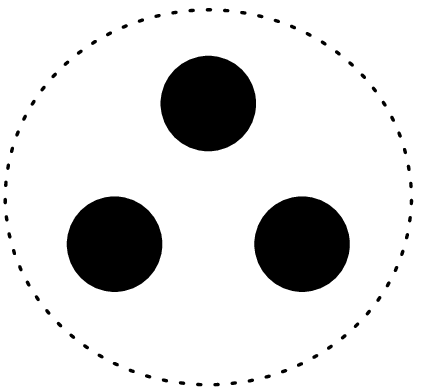}
\savebild{\GD}{\bildGD}{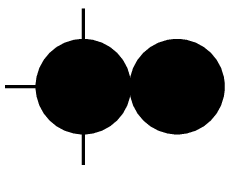}

\setcounter{footnote}{0}

\section{Introduction}\label{introduction}
Interacting lines and more generally manifolds play an important role
in many areas of modern physics. Examples of lines are self-avoiding
polymers \cite{Schaefer}, vortex-lines in super-conductors
\cite{BlatterFeigelmanGeshkenbeinLarkinVinokur1994}, directed polymers
in a disordered environment, also equivalent to surface growth
\cite{KPZ}, diffusion of particles and many more. Generalizing results
to membranes often poses severe problems, but also new insight into
physics. Recently, a lot of work has been devoted to self-avoiding
membranes  (see \cite{WieseHabil} for a review).
Applications reach as far
as high-energy physics, where strings and M-branes have been proposed
as a general framework for unifying all fundamental interactions.
\begin{figure}[b]
  \begin{center}
    \leavevmode
    \raisebox{0mm}{\includegraphics[scale=0.45]{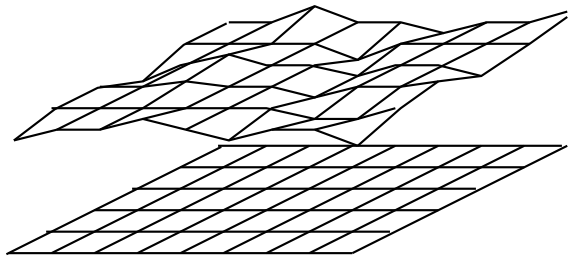}} 
\end{center}
    \caption{A fluctuating membrane interacting by excluded volume with a
    surface-like defect.}
    \label{f1.0}
\end{figure}  

In this article, we focus on the statistical physics of
systems without disorder. We  think of situations like the binding and
unbinding
of a long chain as e.g.\ a polymer  or a membrane on a
wall or the wetting of an interface. More precisely, we study 
 the interaction of a single freely
fluctuating manifold with another non-fluctuating, fixed object. 
Depending on whether the interaction is attractive or repulsive, one
can distinguish two different scenarios: One may either observe a
delocalization transition 
 from an
attractive substrate as in wetting phenomena  or
steric repulsions by a fluctuating manifold. Both cases have in common
that excluded volume effects become important. 
The situation for polymers or a 1-dimensional interface is
 relatively simple: One knows that in this case polymers interacting with a
defect or a wall (excluding or not half of the space) as well as short-range critical wetting
are in the same universality class \cite{Abraham1980,Upton1999}.
For two-dimensional interfaces, the situation is more complicated,
and  a lot of effort has been spent to understand e.g.\ the delocalization
transition
\cite{NakanishiFisher1982,BrezinHalperinLeibler1983,LipowskyKrollZia1983,KrollLipowskyZia1985,FisherHuse1985,LipowskyFisher1987,DavidLeibler1990,ForgasLipowskyNieuwenhuizenInDombGreen}.
Particularly, one is interested in critical wetting for the case of a
short-ranged interaction potential. Then, it can not be approximated by a
polynomial in the field $\vec r$ and the conventional
field-theoretic approach known from $\phi^{4}-$theory fails. This led
to a couple of different ans\"atze 
\cite{BrezinHalperinLeibler1983,LipowskyKrollZia1983,FisherHuse1985,%
LipowskyFisher1987,DavidLeibler1990}, among others the functional
renormalization group approach of \cite{LipowskyFisher1987,DavidLeibler1990}.

Here we follow a different route:  We start by constructing  a
{\em systematic perturbative}  renormalization
group treatment of the delocalization transition as well as the
universal repulsive force exerted by a membrane on a point, line or
more generally hyper-plane like defect.
We therefore introduce a 
flexible ``phantom'' manifold of internal dimension $0\leq D \leq 2$.
By introducing a 
$\delta $-potential  in a  subspace $\mathcal{E}$
of dimension $d'$,  part of the
embedding space $\R^{d+d'}$  (see fig.\ \ref{f1.0}) we punish 
configurations crossing $\mathcal{E}$. Neglecting the
effect of self-avoidance between distinct 
points within the fluctuating manifold, the free energy of a given
configuration   reads:
\begin{equation}
  \label{1.1}
  \mathcal{H}[\vec r] = \int_{x\in\mathcal{M}}\rmd ^{D}x\,\frac{1}{2}(\nabla \vec
  r(x))^{2}+g_{0}\int_{x\in \mathcal{M}}\rmd ^{D}x\int_{\vec{y} \in
\mathcal{E}}\rmd ^{d'}\!\vec{y}\ \, \delta^{d+d'}(\vec
  r(x){-}\vec y)\ ,
\end{equation}
where $\vec{r} (x) $ is the position of monomer $x\in \mathcal{M}$ in
embedding space  $\R^{d+d'}$,
and $\mathcal{M}$ denotes the $D$-dimensional coordinate space of the
manifold. In the case of 
a polymer this is simply the internal chain length. $g_{0}$ is the
attractive or repulsive interaction, of dimension (in inverse length-units)
\begin{equation}
\E (D,d):= \left[g_{0} \right] = D - \frac{2-D}{2} d\ .
\end{equation}
The interaction is naively relevant in the infrared for $\E$ positive, and irrelevant
for $\E$ negative. In a perturbative treatment, $\E$ is the natural
expansion parameter. The situation is similar to self-avoiding
membranes and thus (\ref{1.1}) has been studied as
 a ``toy-model'' for the analysis of
the renormalizability of the more complicated interaction in the generalized
Edwards model for self-avoiding
polymerized membranes \cite{Duplantier1989,DDG1,DDG2}. It was shown to
be renormalizable for arbitrary 
manifold dimensions $0{<}D{<}2$ \cite{DDG1,DDG2}. This means that the
large scale properties are universal, i.e.\ do neither depend on the
regularization scheme used in these calculations, nor on the form of
the contact interaction, as long as it is short-ranged. 
Universal quantities have thus  been obtained to
one-loop order. They are related, as we will show below, to the correction
to scaling exponent $\omega$, which as usually in critical phenomena contains
deviations from the long-distance scaling behavior. 
Neglecting  self-interactions, one has to distinguish between 
 polymers, which are always crumpled  and  membranes, for which  a
highly folded high-temperature state is 
 separated through a crumpling transition
from an essentially flat low-temperature phase
\cite{KantorNelson1987a,KantorNelson1987b,KantorKardarNelson1986a,KantorKardarNelson1986b}.
The effect of self-avoidance can also  be taken into account 
\cite{KardarNelson1987,AronovitzLubensky1988,DDG3,DDG4,WieseHabil}. It
has been shown to lead as in the case 
of polymers to a swelling of the membrane and correspondingly to a
non-trivial exponent for the radius of gyration 
\cite{DavidWiese1996,WieseDavid1997}. However, it is not clear whether
a fractal phase can be found in experiments or simulations. (See
\cite{BowickTravesset2001} for the latest simulations). In this
work  we will mostly neglect self-avoidance. 

The aim of this article is two-fold: First, we present the necessary
techniques to treat the model (\ref{1.1}) beyond the leading order.
We  explicitly perform a two-loop calculation which gives the
correction to scaling exponent $\omega $ at order $\E^2$. Specializing
to membranes one finds that the 2-loop result naively diverges in the limit 
of $D\to 2$. This is a problem of the 
$\E$-expansion,  since there, diagrams have to be evaluated at $\E
(D,d_{c} (D)) =0$, and taking  $D \to 2$ implies 
$d_{c}(D)\to \infty$  causing the result to diverge. 
This motivated us to try a ``massive scheme'' in
fixed dimension, i.e.\ finite $\varepsilon$. It turns out that the limit
$D \to 2$ can then be taken and is regular. Even more, the $2$-loop diagram,
which for $D < 2$ can only be calculated numerically, can now be evaluated
{\em analytically}. This striking property even  holds for higher
orders, and we are 
able to give an explicit -- quite simple -- expression for the perturbation
series. Then  the whole series can be summed and the strong
coupling limit  analyzed. This is one of the very few cases where one
can  indeed obtain the {\em exact} relation
between bare and renormalized coupling in the limit of $D\to2$, and
thus the {\em exact} $\beta$-function.  This result does not depend on the
explicit regularization and renormalization prescriptions, and is also 
obtained for a membrane of spherical or toroidal topology. 
In a final step, we lay the foundations for an 
expansion about $D=2$.  In contrast to the leading order, the first
order corrections already depend on the cut-off procedure. We study 
one specific procedure, which turns out to  
reproduce results for  polymers at leading order approximately, and
even exactly in $d=0$. Work is in progress to obtain a more systematic
expansion about $D=2$ \cite{PinnowWieseProgress}.

\section{Model and physical observables}
\label{Model and physical observables}
\subsection{The model}
We start from the manifold Hamiltonian (\ref{1.1}). We split
the total embedding space $\R^{d+d'}$ into $\mathcal{E}$ and its
orthogonal complement 
$\mathcal{E}_{\perp}$ of dimension $d$. Each $x\in
\mathcal{V}$ points to a point $\vec r(x){=}(\vec r_{\perp}(x),\vec
r_{\parallel}(x))$, with $\vec r_{\parallel}(x)\in \mathcal{E}$ and
$\vec r_{\perp}(x)\in \mathcal{E}_{\perp}$.  The integration over
the subspace $\mathcal{E}$ is then trivial and gives
\begin{equation}
  \label{2.0}
  \int_{\vec{y} \in \mathcal{E}}\delta^{d+d'}(\vec
  r(x){-}\vec y)=\delta^{d}(\vec r_{\perp}(x))\ .
\end{equation}
In the partition function
\begin{equation}
{\cal  Z}_{\ind{total}}  = \int {\cal D}\left[\vec r \right] \, \exp (-{\cal H}\left[\vec r \right])
\end{equation}
the contributions from the parallel and orthogonal components of
$\vec r(x)$  factorize as
\begin{equation}
{\cal Z}_{\ind{total}}  = {\cal Z}_0 \times {\cal Z} (g_{0})
\end{equation}
with
\begin{eqnarray}
  {\cal Z}_0 &=& \int \mathcal{D}[\vec r_{\parallel }]\ \exp\left(-\half
(\nabla_{\parallel }\vec{ r} (x))^{2} \right) \\
\label{2.1a}
\mathcal{Z}(g_{0})&=&\int \mathcal{D}[\vec r_{\perp }]\ \exp(-\mathcal{H}_{\perp}[\vec r_{\perp}])\\
  \label{2.1}
  \mathcal{H}_{\perp}[\vec r_{\perp}]&=& \int_{\mathcal{M}}\ \mbox{d}^{D}\! x \left( \frac{1}{2}(\nabla
 \vec r_{\perp}( x))^{2}+g_{0}\ \delta^{d}( \vec r_{\perp}( x)) \right)\ .
\end{eqnarray}
Since ${\cal Z}_{0}$ is trivial, we will only consider 
$\mathcal{H}_{\perp}[\vec r_{\perp}]$ and shall drop the subscript  
 $_{\perp}$ for notational simplicity. We keep in mind 
that cases with $d<D$ make sense, for instance a polymerized (non-selfavoiding)
membrane interacting with a wall is described by (\ref{2.1})
setting $D=2$ and $d=1$.

\begin{figure}[t]
  \begin{center}
    \leavevmode
    \raisebox{0mm}{\includegraphics[scale=0.45]{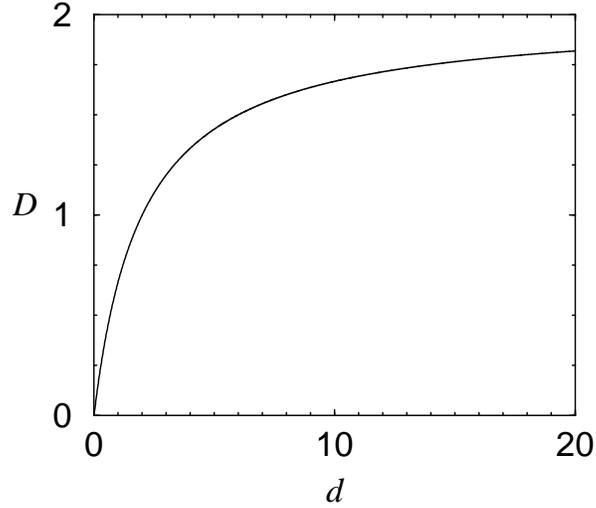}}  
\end{center}
    \caption{Critical line defined through $\varepsilon=0\Leftrightarrow
    d_{c}(D)=\frac{2D}{2{-}D}$. The interaction is relevant for points
that lie above that line.}
    \label{f2.0}
\end{figure}

Let us discuss (\ref{2.1}) in more detail:
The first term is the elastic energy of the manifold which is entropic in
origin. We have scaled elasticity and temperature to unity.
The second term
models the interaction of the manifold with a 
single point at the origin
 in the external $d$-dimensional space, but we remind that the
physical interpretation may well be that of a line or surface.
 The coupling constant $g_{0}$ may either be positive (repulsive
interaction) or negative (attractive interaction).
We now give the dimensional analysis. 
In  coordinate-space units, the engineering dimensions are
\begin{eqnarray*}
  &&\mbox{dim}[x]=1  \\
  \nu&:=&\mbox{dim}[\vec r]=\frac{2-D}{2}
\end{eqnarray*}
\begin{equation}
  \label{2.2}
\varepsilon :=\mbox{dim}\left[ \int_{{\mathcal{M}}}\mbox{d}^{D}\!
  x\ \delta^{d}(\vec r(x))\right] =D-\nu d
\end{equation}
\[g_{0}\ \sim \ \mu^{\varepsilon} \]
\[ \mbox{dim}[\mu]\ =\ -1 \ ,\]\\
where 
\begin{equation}\label{mufromL}
\mu\equiv \frac{1}{L}
\end{equation}
is an inverse length scale. The interaction is naively
relevant for $\E>0$, i.e. $d<d_{c}$ with (see figure \ref{f2.0})
\begin{equation}
  \label{2.3}
  d_{c}=\frac{2D}{2{-}D}\ ,
\end{equation}
irrelevant for $\E<0$ and marginal for $\E=0$.
It has been  shown \cite{DDG1,DDG2} that the model is
renormalizable for $0<D<2$ and $\E\ge0$. Results for negative $\E$ 
are obtained via analytical continuation. 
\begin{figure}[b]
  \begin{center}
    \leavevmode
    \raisebox{0mm}{\includegraphics[scale=0.35]{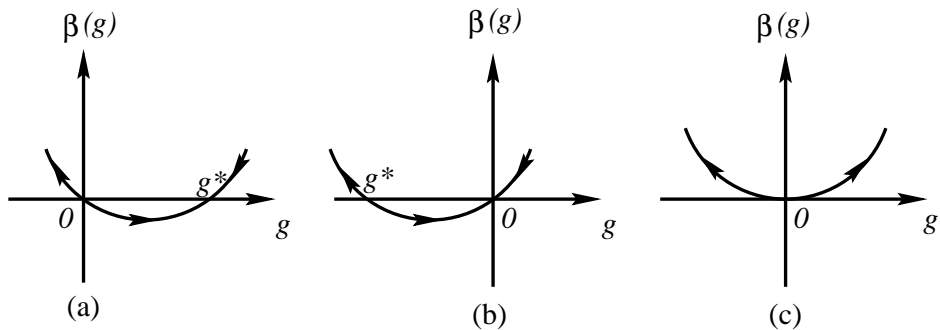}}
  \end{center} 
    \caption{RG-function and flow for increasing manifold size $L$ for
the dimensionless renormalized coupling $g$: (a) in the case
$\varepsilon{>}0$, (b) in the case $\varepsilon{<}0$, (c) in the case
$\varepsilon=0$. }
    \label{f2.1}
\end{figure}
One can define the renormalized coupling $g$ as
\begin{equation}
  \label{2.3.1}
  g:=\frac{\cal N}{\mathcal{V}_{\mathcal{M}}}\left[\mathcal{Z} (0) -\mathcal{Z} (g_{0})\right]L^{\E}\ ,
\end{equation}
where the normalization $\cal N$ depends on the definition of the
path-integral (but not on $L$)  and is
chosen such that 
\begin{equation}\label{2.11p}
g = g_{0}L^{\E}+ O (g_0^{2})\ . 
\end{equation} Universal quantities are obtained
at fixed-points of the $\beta$-function, which is defined as 
\begin{equation}
  \label{2.4}
  \beta(g):=\mu \left.\frac{\partial g}{\partial \mu}\right|_{g_0}\ .
\end{equation}
The $\beta$-function thus  describes, how the effective coupling $g$
changes under scale 
transformations, while  keeping the bare coupling $g_{0}$ fixed. 
Let us already anticipate the 1-loop result, which we derive later. 
It reads
\begin{equation}
  \label{2.5}
  \beta(g)=-\varepsilon g+\frac{1}{2}g^{2}+O(g^{3})\ ,
\end{equation}
where $g$ is the dimensionless renormalized coupling. Apart from the
trivial solution, $g=0$, the flow equation given by (\ref{2.4}) and
(\ref{2.5}) has a  non-trivial fixed point at the zero of the
$\beta$-function
\begin{equation}
  \label{2.6}
  g^{*}=2\varepsilon+O(\varepsilon^{2})\ .
\end{equation}
We shall show below that the scaling behavior is
described by the  slope of the RG-function at the fixed
point, which is universal as a consequence of renormalizabilty.
The long-distance behavior is then governed by the
$\delta$-interaction as considered in our model (\ref{2.1}), which is
the most relevant operator at large scales.  
Let us now discuss  possible physical situations (see fig.\ \ref{f2.1}):
\begin{itemize}
\item [(a)] $\varepsilon{>}0$: The RG-flow has an infrared stable
fixed point at $g^{*}>0$ and an IR-unstable fixed point at $g=0$. The
latter describes an unbinding transition whose critical properties are
given by the non-interacting system, while the non-trivial IR stable
fixed point determines the long-distance properties of the delocalized
state, the long-range repulsive force exerted by the fluctuating
manifold on the origin -- which we remind may be a point, a line or a
plane.
\item [(b)] $\varepsilon{<}0$: Now, the long-distance behavior is
Gaussian, while the unbinding transition occurs at some finite value
of the 
attractive potential,  $g^{*}{<}0$, which corresponds to an infrared
unstable fixed point of the $\beta $-function.  Below $g^{*}$ the
manifold stays always attracted.
\item [(c)] $\varepsilon=0$: This is the marginal situation, where the
transition takes place at $g^{*}=0$; we expect logarithmic corrections
to scaling.
\end{itemize}
Note that in the presence of an impenetrable wall constraining the
configurational space strictly to half of the embedding space, the above considerations should
still apply, when shifting the interaction strength
appropriately. We shall discuss that in \ref{Unbinding transition}. 

\subsection{Repulsive force exerted by a membrane on a wall}\label{wall-force}
Since we are mainly interested in the long-distance properties of membranes
for which always $\varepsilon>0$ (this is (a) on fig.\ \ref{f2.1}),
let us try to calculate the repulsive force exerted by the membrane
on the origin in the case that this point is strictly forbidden. We will
derive a universal expression for this force \cite{DDG2}. 
We need the (not normalized) membrane density at position $\vec{r}$
\begin{equation}\label{2.8}
  \mathcal{Z}(\vec r,g_{0}) :=
\frac{1}{\mathcal{V}_{\mathcal{M}}}\int \mathcal{D}[\vec r]
  \int_{\mathcal{M}}\rmd^{D}x\ \delta^{d}(\vec
r(x)-\vec{r} )\exp[-\mathcal{H}]\ .
\end{equation}
Since the $\delta$-interaction also appears in $\cal H$, we can 
relate the density at the origin to the derivative of the partition
function with respect to $g_{0}$: 
\begin{equation}
  \label{2.9} 
  \mathcal{Z}(\vec 0,g_{0})=- \left. {}\frac{1}{\mathcal{V}_{\mathcal{M}}}
\frac{\partial}{\partial
  g_{0}}\right|_{L}\mathcal{Z}(g_{0})=
\frac{1}{\mathcal{N}} \left.\frac{\partial
(g L^{-\varepsilon})}{\partial 
  g_{0}}\right|_{L}\ ,
\end{equation}
where $g$ is the renormalized or effective coupling defined in
(\ref{2.3.1}). 
Since $\mathcal{Z}(\vec 0,g_{0})$  is dimensionless, it depends 
on the dimensionless combination $g_{0} L^{\E}$ (and
$\varepsilon$) only. Using $\mu =1/L$,  we obtain
\begin{equation}
  \label{2.10}
{\mathcal{N}}  \mathcal{Z}(\vec 0,g_{0})
=\left. L ^{-\E} 
  \frac{\partial g}{\partial  g_{0}}\right|_L
\equiv  -\frac{\beta(g)}{\varepsilon g_{0} L^{\E}}\ .
\end{equation}
For the further considerations, it is sufficient to know the behavior
of the $\beta $-function close to the nontrivial zero. Expanding about
$g{=}g^{*}$, we have
\begin{equation}
  \label{2.11}
  \beta(g) = (g-g^{*})\omega(g^{*})+O ((g-g^{*})^{2})\ .
\end{equation}
Combining (\ref{2.9}), (\ref{2.10}) and (\ref{2.11}), we get
\begin{equation}
\left.\frac{\partial g }{\partial
  ( g_{0} L^{\varepsilon})}\right|_{L} =  -\frac{\beta(g)}{\varepsilon g_{0} L^{\E}}
=-\frac{(g-g^{*})\omega(g^{*}) }{\varepsilon g_{0} L^{\E}} \ .
\end{equation}
The solution of this differential equation is
\begin{equation}
  \label{2.12}
  g-g^{*}\sim  (g_{0}L^{\E})^{-\omega(g^{*})/\varepsilon} + \dots \ 
\ .
\end{equation} 
Using  (\ref{2.10}), this implies  the scaling law
\begin{equation}
  \label{2.13}
  \mathcal{Z}(\vec 0,g_{0})\sim (g_{0}^{1/\varepsilon}L)^{-\omega(g^{*})-\varepsilon}\ .
\end{equation}
We now use this result to derive a scaling law for $\mathcal{Z}(\vec
r,g_{0})$, with $\vec{r} \neq \vec{0} $. In addition to $g_{0}L^{\E}$
(and $\E$), $\mathcal{Z}(\vec r,g_{0})$ depends on $\vec{r} $; taking
into account rotational symmetry and dimensionality, it depends
on $\vec r$ only through the dimensionless combination $r/L^{\nu }$, with $r=|\vec{r}| $:
\begin{equation}
  \label{2.16}
  \mathcal{Z}(\vec r,g_{0}) = 
  \mathcal{Z}(r/L^{\nu},g_{0}L^{\E})\ .
\end{equation}
For  $\mathcal{Z}(\vec r,g_{0})$, the most interesting limit is that
of $g_{0}\to \infty $. Physically, this corresponds to strictly forbidding
monomers to be at the origin. Therefore it is clear that this limit is
well-behaved, and  
\begin{equation}
  \label{2.17} \mathcal{Z}_{\infty}(r/L^{\nu}):=
  \lim_{g_{0}\to \infty}\mathcal{Z}(r/L^{\nu},g_{0}L^{\E})
\end{equation}
is finite. In order to be consistent with (\ref{2.13}) it has to 
obey a power law in the scaling regime $r\ll L^{\nu}$
\begin{equation}
  \label{2.18}
  \mathcal{Z}_{\infty}(r/L^{\nu})\sim (r/L^{\nu})^{\theta}\ .
\end{equation}
Comparing the $L$-dependence of $\mathcal{Z}_{\infty } (r/L^{\nu })$
and $\mathcal{Z} (\vec{0},g_{0} )$, we obtain the  exponent identity
\begin{equation}
  \label{2.19}
  \theta=\frac{\varepsilon+\omega(g^{*})}{\nu}\ .
\end{equation}
Finally, from (\ref{2.18}) we derive the repulsive force between the
origin and the manifold 
\begin{equation}
  \label{2.20}
  \vec f(\vec r)= \nabla_{\vec r}\ln \mathcal{Z}_{\infty }(| \vec{r}
|/L^{\nu }) = \theta \frac{\vec r}{r^{2}}\ .
\end{equation}
Note that to derive this result, $k_{B}T$
has been set to 1. Reestablishing the temperature-dependence, we find
\begin{equation}
  \vec f(\vec r)= k_{B}T \theta \frac{\vec r}{r^{2}}\ .
\end{equation}
Also note that this argument gives $\theta =0$ at the
Gaussian fixed point, which is necessary since for $g_{0}=0$ no force
is excerted on the membrane.

\subsection{Unbinding transition}\label{Unbinding transition}
Let us discuss the physical situation at the UV-stable fixed point in figure
\ref{f2.1}. The fixed point corresponds to a {\em delocalization
transition} of the manifold, which is at vanishing coupling $g^{*}{=}0$ for $\varepsilon{>}0$
and  at some finite attractive coupling $g^{*}{<}0$ for
$\varepsilon{<}0$.\\
In the localized phase $g{<}g^{*}$, correlation functions such as
$\left<[\vec r(x)-\vec r(y)]^{2}\right>$ and the associated correlation length
$\xi_{\parallel}$ (in the $D$-dimensional internal space)  should be finite, as
well as the radius of gyration $\xi_{\perp}$. Approaching the transition point
 these quantities diverge as \cite{ForgasLipowskyNieuwenhuizenInDombGreen}
\begin{equation}
  \label{u.1}
  \xi_{\parallel}\ \sim \ (g^{*}{-}g)^{-\nu_{\parallel}}\quad ,\quad
  \xi_{\perp}\ \sim \ (g^{*}{-}g)^{-\nu_{\perp}}\ .
\end{equation}
Since $\xi_\perp\sim\xi_\parallel^{\nu }$, the exponents $\nu_{\parallel}$ and $\nu_{\perp}$ are related through
\begin{equation}
  \label{u.2}
  \nu_{\perp}\ =\ \nu_{\parallel}\nu\ ,
\end{equation}
$\nu$ being the dimension of the field (\ref{2.2}).
To relate $\nu_{\parallel }$ to $\omega $, we first observe that $\xi
_{\parallel }$ depends          only on $g_0$
and the membrane size $L$, or its inverse $\mu =1/L$.  Writing $\xi
_{\parallel }$ as a function of the renormalized coupling $g$ and the
 scale $\mu $, we thus have 
\begin{eqnarray}
0&=&\mu          \frac{\partial }{\partial \mu } \xi _{\parallel } (g,\mu ) \nn\\&=&
\mu      \frac{\partial }{\partial \mu } \left(\frac{1}{\mu }f (g)
\right)\nonumber  \\
&=&  \frac{1}{\mu }\left(-f (g) +\beta (g)f' (g) \right)\ ,
\end{eqnarray}
where the dimension-full factor  $1/\mu $ has been factored from the
dependence on the dimensionless renormalized coupling $g$.
 Using that $\frac{\partial }{\partial g}\xi _{\parallel }
(g)=\frac{1}{\mu }f' (g)$, this can be rewritten as
\begin{eqnarray}
\xi _{\parallel } (g) &=& \beta (g)\frac{\partial }{\partial g}\xi
_{\parallel } (g)\nn\\
&\approx & \omega (g^{*}) (g-g^{*})\frac{\partial }{\partial g}\xi
_{\parallel } (g)
\ .
\end{eqnarray}
The solution to the above equation is 
\begin{equation}
\xi _{\parallel } \sim  |g-g^{*}| ^{1/\omega (g^{*})}\ .
\end{equation}
This leads to the identification 
\begin{equation}
\nu _{\parallel } = -\frac{1}{\displaystyle \omega (g^{*})} \ , \qquad
\nu _{\perp  } = -\frac{\nu }{\displaystyle \omega (g^{*})} \ .
\end{equation}
Note that $\omega(g^{*}){<}0$ at the transition. Specializing to
$(D,d){=}(1,1)$, we find
\begin{equation}
  \label{u.6}
  \nu_{\perp}\ =\ 1\quad ,\quad \nu_{\parallel}\ = \ 2\ .
\end{equation}

These exponents are also valid for  the delocalization transition of a
$1$-dimensional interface from an attractive hard wall in $2$-dimensional bulk
space \cite{BrezinHalperinLeibler1983,ForgasLipowskyNieuwenhuizenInDombGreen,Upton1999}. This can be understood as follows: The  partition function of a
fluctuating polymer interacting with a $\delta$-defect at the
boundary of a hard wall can be written as
\begin{eqnarray}
  \label{u.7}
  \mathcal{Z}\ =\
  \frac{1}{2}\sum_{n=0}^{\infty}\mathcal{Z}_{n}\left(\frac{1}{2}\right)^{n}\
  e^{-n\beta E}\ ,
\end{eqnarray}
\begin{figure}
$\parbox{0.45\textwidth}{\epsfxsize=0.45\textwidth\epsfbox{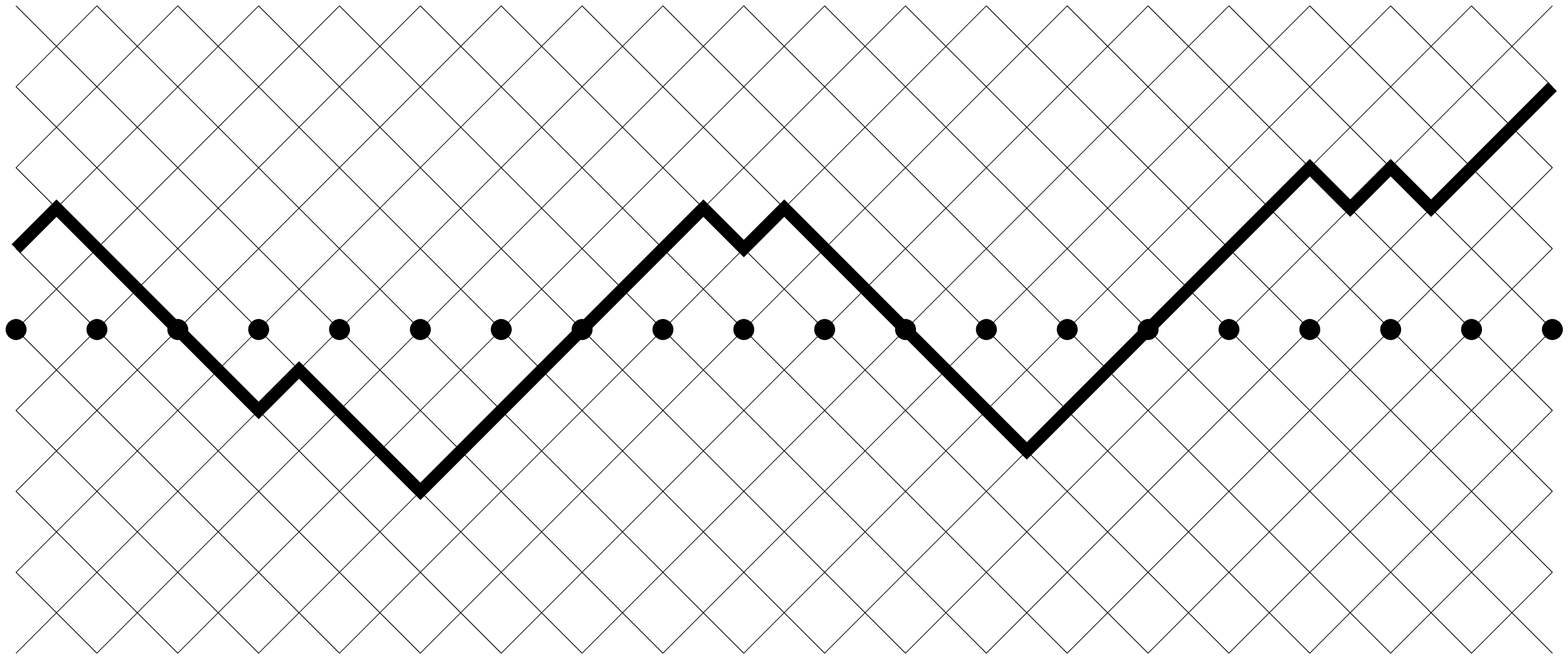}}
\hfill \longrightarrow \hfill 
\parbox{0.45\textwidth}{\epsfxsize=0.45\textwidth\epsfbox{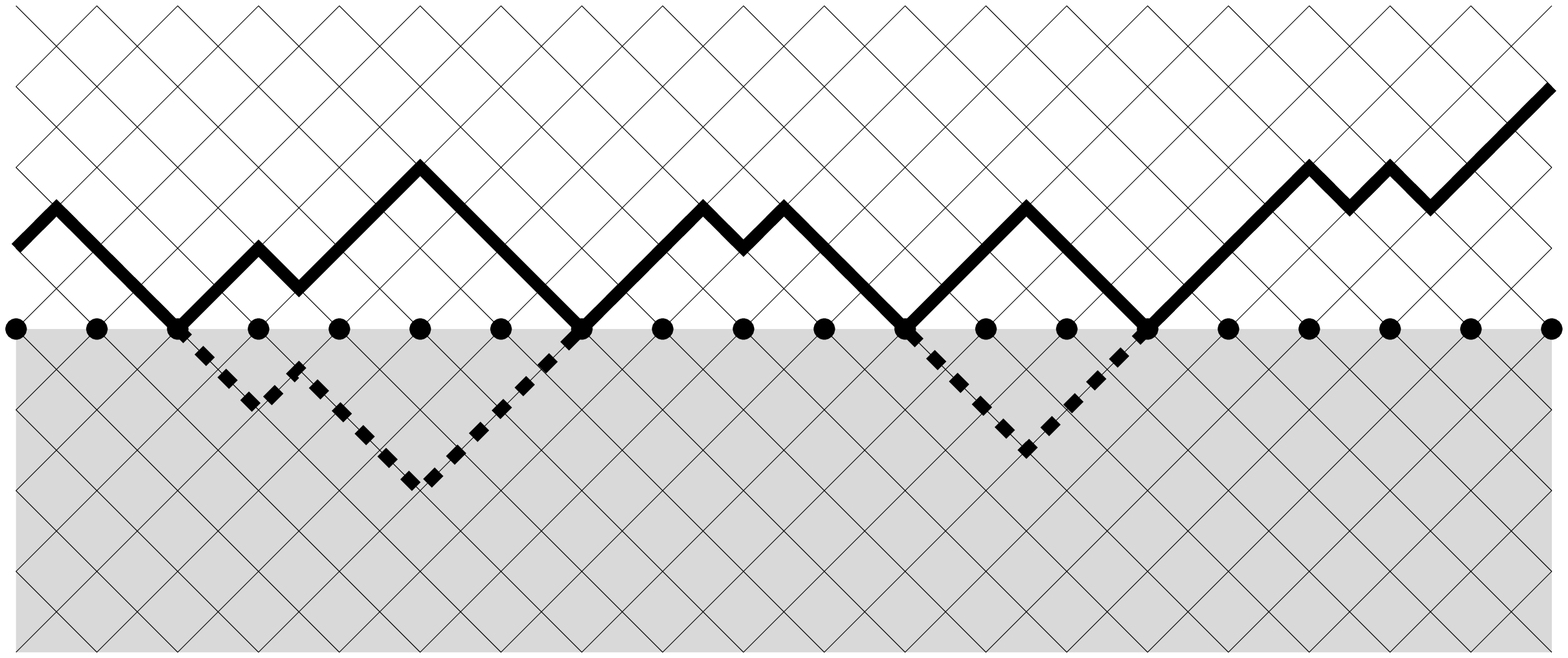}}$
\caption{A polymer (here for simplicity directed) in interaction with
a wall (thick dots). Left before and right after flapping up. The grey
area is impenetrable.}
\label{flap}
\end{figure}
where the sum runs over the number of contacts with the defect hyper-plane. $E$
denotes the contact energy, $\beta=(k_{B}T)^{-1}$ and $\mathcal{Z}_{n}$
is the constrained partition function of the free interface having exactly
$n$ points of contact with the defect. The powers of $\frac{1}{2}$  only
appear in the presence of an impenetrable defect and reflect the fact
that all configurations in the presence of a hard wall can be obtained
by flapping up those parts of the polymer which have penetrated the
line (see figure \ref{flap}). This accounts for a factor $\frac{1}{2}$
for each possible flap.
 Thus, the hard wall
constraint translates  into a shift in the binding energy according to 
\begin{equation}
  \label{u.8}
  E_{\mathrm{no\;wall}}\ \longrightarrow\ E_{\mathrm{wall}} =
  E_{\mathrm{no\;wall}}+k_{B}T\ \ln 2\ .
\end{equation}
The delocalization transition occurring for $(D,d){=}(1,1)$ at vanishing
potential strength is shifted to an attractive interaction dependent on the
temperature. Due to the correspondence (\ref{u.8}) the exponents
characterizing the transition remain unchanged.


\section{Operator product expansion and one-loop result}
The partition function of the model has been  defined in (\ref{2.1a}).
Analogously, we define the expectation value of an observable $
\mathcal{O}(r)$ as (we denote $r$ for $\vec{r} $ from now on)
\begin{equation}
  \label{3.1a}
\left< \mathcal{O}(r)   \right>_{g_{0}}  :=\frac{\int \mathcal{D}[r]\ \exp(-\mathcal{H}[r])}{Z (g_{0})}\ .
\end{equation}
Most physical observables can be derived from expectation values of general
$M$-point vertex operators like
\begin{equation}
  \label{3.2}
  \mathcal{O}(r)=\prod_{j=1}^{M}\rme^{ip_{j}r(s_{j})}\ .
\end{equation}
Before we go on, let us make some changes in normalizations which will
be helpful in the following. First, we rescale the field and the coupling constant according to
\begin{equation}
  \label{3.8}
  r \longrightarrow (2{-}D)^{1/2}r\ ,\quad g_{0} \longrightarrow
 (2{-}D)^{d/2}g_{0}\ .
\end{equation}
Second, we change the integration over internal coordinates to 
\begin{equation}
  \label{3.8.1}
  \int_{x}:=\frac{1}{S_{D}}\int \mbox{d}^{D}x\ ,\quad S_{D}=2\frac{\pi^{D/2}}{\Gamma(D/2)}
\end{equation}
being the volume of the $D$-dimensional unit-sphere. Third, the
normalization of the $\delta$-distribution is changed to 
\begin{equation}
  \label{3.8.2}
  \tilde \delta^{d}(r(x)):=(4\pi)^{d/2}\delta(r(x))=\int_{k}\rme^{ikr(x)}
\end{equation}
with
\begin{equation}
  \label{3.8.3}
  \int_{k}:=\pi^{-d/2}\int \mbox{d}^{d}k
\ .
\end{equation}
The  advantage of these normalizations is that 
\begin{equation}
\int_{k} \rme^{-k^{2}} = 1\ , \quad \int_{x} \Theta (|x|<1) =1 \ .
\end{equation}
The model in the new normalizations now reads
\begin{equation}\label{newmodel}
{\cal H}[r] = \frac{1}{2-D}\int_{x}\half \left(\nabla r (x)  \right)^{2}+
g_{0} \int_{x}\tilde{\delta }^{d} (r (x))\ ,
\end{equation}
and (due to the factor of $1/ (2-D)$ in the above definition) the
two-point correlator is 
\begin{equation}
  \label{3.7}
  C(x_{i}{-}x_{j}):=\frac{1}{2d}\left<[r(x_{i})-r(x_{j})]^{2}\right>_{0}=|x_{i}-x_{j}|^{2\nu}
\ .
\end{equation}

We now proceed with the calculation of physical observables. 
As an explicit example, let us consider the perturbation expansion of the
$1$-point vertex operator
\begin{equation}
  \label{3.3}
  \mathcal{Z}^{(p)}:=\mathcal{Z} (g_{0}) \langle \rme^{ipr(s)}\rangle_{g_{0}}=\sum_{N=0}^{\infty}\frac{(-g_{0})^{N}}{N!}\mathcal{Z}^{(p)}_{N},
\ ,
\end{equation}
where
\begin{equation}
  \label{3.4}
  \mathcal{Z}^{(p)}_{N}=\EXP{ \rme^{ipr(s)} \prod^{N}_{i=1}\int_{x_{i}}\
  \tilde\delta^{d}(r(x_{i}))}{\ 0}\ .
\end{equation}
This can be written as
\begin{equation}
  \label{3.5}
  \mathcal{Z}^{(p)}_{N}=
  \prod_{i=1}^{N}\int_{x_{i}}\int_{k_{i}}\tilde\delta^{d}(p+\sum
k_{i})\times \EXP{ \exp\left[ i ( \sum_{i=1}^{N}k_{i}r
  (x_{i})+pr(s))\right]}{\ 0}\ ,
\end{equation}
where we have already integrated out a global translation of the field $r$.
The Gaussian average is:
\begin{equation}
  \label{3.6}
    \EXP{\exp \left[i(\sum_{i=1}^{N}k_{i}{r}
  (x_{i})+pr(s))
  \right]}{\ 0}=\exp\left[\frac{1}{2}\sum_{i,j=0}^{N}k_{i}k_{j}C(x_{i}{-}x_{j})\right]\ \
  ,\ k_{0}=p
\ .
\end{equation}
The $\mathcal{Z}^{(p)}_{N}$ in (\ref{3.4}) posses short distance
singularities for $N{\geq}2$. In order to analyze these singularities we will
use  the techniques of normal ordering and operator product
expansion in the sequel.
In our problem the procedure of normal ordering a vertex operator
$\rme^{ik_{j}r(x_{j})}$ turns out to be quite simple:  In any
expectation value, we factorize out the  contractions between the 
 vertex operators. At one-loop order or
equivalently at second order in (\ref{3.3}) this means that
\begin{equation}
  \label{3.9}
  \rme^{ik_{1}r(x_{1})}\rme^{ik_{2}r(x_{2})}\ =\
  \bn \rme^{i(k_{1}r(x_{1})+k_{2}r(x_{2}))}\en\ \rme^{k_{1}k_{2}C(x_{1}{-}x_{2})}\ .
\end{equation}
This can be understood as a definition of the normal ordered product $\bn
\rme^{i(k_{1}r(x_{1})+k_{2}r(x_{2}))}\en$. Note, that equality in the above equation is meant in the sense of
operators, that is when inserted into expectation values within the free
theory.\\
Let us now turn to an explicit derivation of the OPE of two
$\delta$-interactions, i.e. we study (\ref{3.9}) for small distances
$x_{1}{-}x_{2}$. Since the short-distance singularities appear only in the
internal contractions, the normal ordered product in the r.h.s. of (\ref{3.9})
is regular for small distances $x_{1}{-}x_{2}$ and thus can be expanded
therein. We then project the resulting terms on the corresponding 
operators. For that purpose we make a change in the momentum variables
according to
\begin{equation}
  \label{3.10}
  \left(
  \begin{array}[h]{c}
    k_{1} \\
    k_{2}
    \end{array}
    \right) \longrightarrow 
    \left(
      \begin{array}[h]{c}
        \frac{k_{1}}{2}-k_{2} \\
        \frac{k_{1}}{2}+k_{2}
      \end{array}
    \right)\ .
\end{equation}
In internal space we change coordinates to the center of mass system:
\begin{equation}
  \label{3.11}
  \bar x:=\frac{1}{2}(x_{1}+x_{2})\ ,\quad
  y:=x_{2}-x_{1}\ .
\end{equation}
Thus, the r.h.s. of (\ref{3.9}) takes the form
\begin{eqnarray}
  \label{3.12}
  &&\bn \rme^{i[(\frac{k_{1}}{2}{-}k_{2})r(\bar
  x{-}y/2){+}(\frac{k_{1}}{2}{+}k_{2})r(\bar x{+}y/2)]}\en\,
\rme^{(k_1^{2}/4-k_{2}^{2})C (y)}
\ =\ \nonumber\\
  &&\bn \rme^{i k_{1}r(\bar x) }\ \left(1+ik_{2}y\nabla
  r(\bar x)+ik_{1}\frac{1}{2!}\frac{y^{\alpha}y^{\beta}}{4}\nabla_{\alpha}\nabla_{\beta}r(\bar x)-\frac{(k_{2}y\nabla r(\bar x))^{2}}{2!}+\dots\right)\en\times\nonumber\\
  &&\hphantom{\bn \rme^{ik_{1}r(\bar x)}\ }\left(1+\frac{k_{1}^{2}}{4}C(y)+\dots\right)\
  \rme^{-k_{2}^{2}C(y)}\ .
\end{eqnarray}
Integration over  $k_{1},k_{2}$ yields
 the OPE of $\tilde\delta^{d}(r(x_{1}))$ with $ \tilde\delta^{d}(r(x_{2}))$:
\begin{eqnarray}
  \label{3.13}
 \tilde \delta^{d}(r(\bar x-\frac{y}{2}))\tilde \delta^{d}(r(\bar x+\frac{y}{2}))
  &=& y^{-\nu d} \tilde \delta^{d}(r(\bar x))-y^{-\nu
d+2\nu}\frac{1}{4}\ \partial_{r}^{2}
\tilde  \delta^{d}(r(\bar x))\nonumber\\
  &&-y^{-\nu d-2\nu+2}\frac{1}{4D}\tilde \delta^{d}(r(\bar x)) (\nabla
  r(\bar x))^{2}\nonumber \\
&& -y^{-\nu d+2}\frac{1}{16D} (-\Delta_{r})\tilde \delta^{d}(r(\bar x))
  (\nabla r(\bar x))^{2}\nonumber\\
  &&+y^{-\nu d+2}\frac{1}{8D}\ \partial_{r_{i}}
\tilde\delta^{d}(r(\bar x))  \Delta
  r_{i}(\bar x)+\dots\  .
\ .
\end{eqnarray}
Restricting ourselves to the most relevant next to leading operators for
internal dimension $D$ smaller than, but  close to $2$, the operator product expansion reads:
\begin{equation}
  \label{3.14}
\tilde\delta^{d}(r(\bar x-\frac{y}{2}))\tilde \delta^{d}(r(\bar x+\frac{y}{2}))\
  =\ y^{-\nu d}\tilde\delta^{d}(r(\bar x))+\sum_{n=1}^{\infty}y^{-\nu
  d+n(D-2)}\frac{(-\partial_{r}^{2})^{n}\tilde\delta^{d}(r(\bar x))}{4^{n}n!}
\ .
\end{equation}
The OPE-coefficient in front of the leading operator, which is the
$\delta$-interaction itself, carries the leading short distance singularity. In the
following we introduce some diagrammatic notation.
 Denoting $\GA :=\tilde\delta^{d}(r(x))$ we abbreviate the projection of two nearby
$\delta$-interactions on the $\delta$-interaction itself as
\begin{equation}
  \label{3.15}
  \MOPE{\GB}{\GA}\ =\
  y^{-\nu d}\ .
\end{equation}
We now analyze short-distance divergences of the perturbation expansion using
the OPE in (\ref{3.13}). Being interested in the leading scaling behavior, we
drop all subdominant terms for $0<D<2$, such that we only need the leading
OPE-coefficient (\ref{3.15}).
The integral over the relative distance in $\int_{y}\MOPE{\GB}{\GA}\GA$ is
logarithmically divergent for $\varepsilon{=}0$. In order to define a counter
term we use dimensional regularization, i.e.\ we set
$\varepsilon{>}0$. The integral is then UV-convergent, but IR-divergent. An IR regulator or equivalently a subtraction scale
$\mu{=}L^{-1}$ has to be introduced to define the subtraction
operation. Generally, we integrate over all distances bound by the
inverse subtraction scale $L$. The OPE-coefficient in (\ref{3.15}) then
reduces to a number as
\begin{equation}
  \label{3.16}
  \DIAGind{\GB}{\GA}{L}:=\int_{|y|{<}L}\MOPE{\GB}{\GA}=L^{\varepsilon}f(\varepsilon,D)\ .
\end{equation}
Let us use the scheme of minimal subtraction (MS). The internal dimension of the
membrane is fixed and (\ref{3.16}) is expanded in a Laurent-series in
$\varepsilon$, starting here at order $\varepsilon^{-1}$. Denoting the term of
order $\varepsilon^{p}$ in $\left.\DIAGind{\quad}{\quad}{L}\right|_{L=1}$ by
$\DIAGind{\quad}{\quad}{\varepsilon^{p}}$, the simple pole in (\ref{3.16}) is
\begin{equation}
  \label{3.17}
  \DIAGind{\GB}{\GA}{\varepsilon^{-1}}=\frac{1}{\varepsilon}\ .
\end{equation}
Of course, in this case, due to our normalizations, this is exact.\\
Let us rewrite the Hamiltonian in (\ref{2.1}) in terms of the renormalized
dimensionless coupling $g$. For perturbative calculations it is custom
to write
\begin{equation}
  \label{3.18}
  g_{0}=gZ_{g}\mu^{\varepsilon}\ ,
\end{equation}
where $Z_{g}$ is the renormalization factor, which is fixed by
demanding that observables remain finite in the limit of $\E\to0$. The
Hamiltonian becomes
\begin{equation}
  \label{3.19}
  \mathcal{H}[r]= \frac{1}{2{-}D}\int_{x}\, \frac{1}{2}(\nabla
 r(x))^{2}+\int_{x} gZ_{g}\mu^{\varepsilon}\,\tilde \delta^{d}(r(x)) \ .
\end{equation}
We find to one-loop order
\begin{equation}
  \label{3.20}
  Z_{g}=1+\frac{g}{2}\DIAG{\GB}{\GA}{_{\varepsilon^{-1}}}+O(g^{2})\ .
\end{equation}
There is no field-renormalization since the elastic part has also to
describe the manifold far away from the origin, for which the
interaction term can be neglected.
 Formally, of course, this follows from the OPE.
The renormalization group $\beta$-function is defined as
\begin{equation}
  \label{3.21}
  \beta(g):=\left.\mu \frac{\mbox{d}}{\mbox{d}\mu}\right|_{g_{0}}g\ ,
\end{equation}
from what follows together with (\ref{3.18}) that
\begin{equation}
  \label{3.22}
  \beta(g)=\frac{-\varepsilon g}{1+g\frac{\partial}{\partial g}\ln Z_{g}}\ .
\end{equation}
To one-loop order we obtain from (\ref{3.22}) and (\ref{3.20}) as
anticipated earlier
\begin{equation}
  \label{3.23}
  \beta(g)=-\varepsilon g+\frac{g^{2}}{2}+O(g^{3})\ .
\end{equation}
The long-distance behavior of the theory is governed by the IR--stable
fixed point of the RG--flow. Fixed points are zeros of the
$\beta$--function:
\begin{equation}
  \label{3.24}
  \beta(g^{*})=0\quad \Rightarrow\quad  g^{*}=0\quad \textrm{or} \quad
  g^{*}=2\varepsilon\ .
\end{equation}
In section \ref{wall-force}, we had defined the correction-to-scaling 
exponent $\omega$. It is obtained from the slope-function $\omega
(g)$, defined as
\begin{equation}
\omega (g):= \frac{\rmd }{\rmd g} \beta (g)\ .
\end{equation}
The correction to scaling exponent is $\omega (g)$ evaluated at the
fixed points, and is found to be 
\begin{equation}
  \label{3.25}
  \omega(g^{*}{=}0)=-\varepsilon\ , \qquad \omega(g^{*}{=}2\varepsilon)=\varepsilon\ .
\end{equation}
Of course, this result  is apart from the very existence of a fixed
point  rather trivial, since it is determined
through the leading 
term of the $\beta$-function, which is always $-\varepsilon g$. In order
to obtain any non-trivial information from the perturbation series
 it is
thus necessary to go beyond the leading order.


\section{Two-loop calculation in a MS-scheme}
\subsection{Operator product expansion}
Let us now continue with the calculation at the two-loop order. At
each order of perturbation theory there is only one new diagram. At
two-loop order this comes from three coalescing
$\delta$-interactions. Again, let us rewrite the product of the three
vertex operators as its expectation value times the normal ordered
product:
\begin{eqnarray}
  \label{4.1}
  &&\rme^{ik_{1}r(x_{1})}\rme^{ik_{2}r(x_{2})}\rme^{ik_{3}r(x_{3})}\nonumber\\
  &&=\
  \bn \rme^{i[k_{1}r(x_{1})+k_{2}r(x_{2})+k_{3}r(x_{3})]}\en \ \rme^{k_{1}k_{2}C(x_{1}{-}x_{2})}\rme^{k_{1}k_{3}C(x_{1}{-}x_{3})}\rme^{k_{2}k_{3}C(x_{2}{-}x_{3})}\ .
\end{eqnarray}
We use the following change of coordinates
\begin{eqnarray}
  \label{4.2}
  \left.
    \begin{array}{r@{\ :=\ }l}
      \bar x & \frac{1}{3}(x_{1}+x_{2}+x_{3}) \\
      a & x_{2}-x_{1} \\  
      b & x_{3}-x_{2} \\
      c & x_{1}-x_{3}
    \end{array}\right\} \Leftrightarrow \left\{ 
    \begin{array}{r@{\ =\ }l}
      x_{1} & \bar x+\frac{1}{3} (c-a)\rule{0mm}{1.ex} \\
      x_{2} & \bar x+\frac{1}{3} (a-b)\rule{0mm}{2.5ex} \\
      x_{3} & \bar x+\frac{1}{3} (b-c)\rule{0mm}{2.5ex} \\
    \end{array}\right. \ ,
\ .
\end{eqnarray}
As at 1-loop order, $ \bn
\rme^{i[k_{1}r(x_{1})+k_{2}r(x_{2})+k_{3}r(x_{3})]}\en$ is free of
divergences upon approaching the points $x_{i}$, whereas the UV-divergence
comes from the factor $\rme^{k_{1}k_{2}C(x_{1}{-}x_{2})}
\rme^{k_{1}k_{3} C(x_{1}{-}x_{3})}\rme^{k_{2}k_{3}C(x_{2}{-}x_{3})}$.
The leading term is thus obtained upon setting $x_{1}=x_{2}=x_{3}=\bar
x$ in  $ \bn \rme^{i[k_{1}r(x_{1})+k_{2}r(x_{2})+k_{3}r(x_{3})]}\en$,
 dropping the gradient terms. Further shifting 
$k_{1}\to k_{1}-k_{2}-k_{3}$, we obtain 
\begin{eqnarray}
\rme^{i k_{1}r (\bar x)} \, \rme^{( k_{1}-k_{2}-k_{3})k_{2}C (a)+k_{2}k_{3}C (b)+ ( k_{1}-k_{2}-k_{3})k_{3}C ( c)}
\ .
\end{eqnarray}
Integrating over $k_{1}$, $k_{2}$ and $k_{3}$ yields the final result
\begin{eqnarray} \label{4.7}
\tilde \delta ^{d} (r (x_{1}))\tilde \delta ^{d} (r (x_{2}))
\tilde \delta ^{d} (r (x_{3}))=
\MOPE{\GC}{\GA}\tilde  \delta^{d}(r(\bar x))+\MOPE{\GC}{\GA^{\prime\prime}}\
    (-\partial_{r}^{2})\tilde \delta^{d}(r(\bar x))+\dots\ ,\nonumber \\
\end{eqnarray}
where the OPE--coefficients are given by
\begin{eqnarray}
  \label{4.8}
  \MOPE{\GC}{\GA}&=&
  \left[C(a)C(c)-\frac{1}{4}(C(a)+C(c)-C(b))^{2} \right]^{-d/2}\nonumber \\
&=&
  \left[\frac{1}{4}(a^{\nu}{+}b^{\nu}{+}c^{\nu})(b^{\nu}{+}c^{\nu}{-}a^{\nu})(a^{\nu}{+}c^{\nu}{-}b^{\nu})(a^{\nu}{+}b^{\nu}{-}c^{\nu})\right]^{{-}d/2}
\ ,
\end{eqnarray}
which contributes to the renormalization factor at two-loop order.
The first subleading term is (denoting $\GA'':=(-\partial_r^{2})\tilde
\delta ^{d} (r (\bar x))$)
\begin{eqnarray}
  \label{4.9}
  \MOPE{\GC}{\GA^{\prime\prime}}=-\frac{1}{4}C (a)C (b)C (c) \left[\frac{1}{4}(a^{\nu}{+}b^{\nu}{+}c^{\nu})(b^{\nu}{+}c^{\nu}{-}a^{\nu})(a^{\nu}{+}c^{\nu}{-}b^{\nu})(a^{\nu}{+}b^{\nu}{-}c^{\nu})\right]^{{-}d/2{-}1}\nonumber \\
\end{eqnarray}
\subsection{Numerical calculation in $0<D<2$}
Let us now turn to the explicit calculation of the second order contribution to
the coupling constant renormalization. To two-loop order the
renormalization group $Z_{g}$-factor for the coupling
constant reads
\begin{eqnarray}
  \label{4.11}
  Z_{g} = 1+ \frac{g}{2}
  \DIAG{\GB}{\GA}{_{\varepsilon^{-1}}}&-&\frac{g^{2}}{6}\left[\DIAG{\GC}{\GA}_{\varepsilon^{-2},
  \varepsilon^{-1}}-\frac32\DIAG{\GB}{\GA}_{\varepsilon^{-1}}^{2}\right]\nonumber\\
  &+&\frac{g^{2}}{4}\DIAG{\GB}{\GA}_{\varepsilon^{-1}}^{2}+O(g^{3})
\ .
\end{eqnarray}
Note that this $Z_{g}$-factor can either be obtained from
(\ref{2.3.1}) and (\ref{3.18}) or by expanding the perturbative series and checking
that divergent contributions proportional to $\tilde \delta (r (x))$
cancel.

The formula is arranged such that in the brackets the second order pole
in $\varepsilon$ cancels. This is, because to leading order in $1/\E$,
the two-loop diagram factorizes into two one-loop diagrams. The
combinatorial factor of $3/2 $ is composed of a factor of 3 for the
number of possible subdivergences, and a factor of $1/2$ for the nested
integration: The subdivergence in the distance say $a$ only appears in
the sector 
 (i.e.\ part of the domain of integration) in which $a$ is  smallest.

Let us define:
\begin{equation}
  \label{4.12}
  I_{2}(D,L) :=\DIAG{\GC}{\GA}_{L}{-}\frac{3}{2}\DIAG{\GB}{\GA}_{L}^{2}\ .
\end{equation}
Let us already note that we will 
later calculate in a ``massive'' scheme in fixed
dimension $d$. In order to use the results derived here, we keep $d$ arbitrary
instead of setting  $d=d_{c}$. We now want to derive an expression
which can be integrated numerically. Since the bounds on the
integrations in (\ref{4.12}) are asymmetric (remind that in
$\DIAG{\GC}{\GA}_{L}$ all 3 distances $a$, $b$, and $c$ are bounded by
$L$,
whereas $ \DIAG{\GB}{\GA}_{L}^{2}$ is an integral with two bounds only)
we treat both terms separately. We start with the first one, which can
be written as 
\begin{eqnarray}
  \label{4.16}
  I_2^{\ind{M}}(D,L)&=&\int_{a,b,c}\Theta(a\leq L)\Theta(b\leq
L)\Theta(c\leq L) \MOPE{\GC}{\GA}
\ .
\end{eqnarray}
The divergence when integrating over the global scale is eliminated 
by noting that $ L\frac{\partial}{\partial L}I_{2}(D,L)=2\varepsilon\
I_{2}(D,L)$. There remains of course the sub-divergences, which will
be subtracted later by the counter-term.
 Applying this procedure to (\ref{4.16}), we obtain
\begin{eqnarray}
  \label{4.17}
  2\E   I_2^{\ind{M}}(D,L)&=& \int_{a,b,c}\Big[\delta(a-L)\Theta(b\leq L)\Theta(c\leq
  L)+\delta(c-L)\Theta(b\leq L)\Theta(a\leq L) \nonumber\\
  && \qquad +\delta(b-L)\Theta(a\leq L)\Theta(c\leq L)\Big] \MOPE{\GC}{\GA}\ .
\end{eqnarray}
Since the integral is symmetric in $a$, $b$, and $c$, we can replace
it by 3 times the first term. Introducing proper normalizations,
an explicit representation for the measure and factoring out the
explicit $L$-dependence, we obtain
\begin{eqnarray}
  \label{4.18}
  2\E   I_2^{\ind{M}}(D,L)&=& 3 L^{2\E}\frac{S_{D-1}}{S_{D}}\int_{-
  \infty}^{\infty}\mbox{d}c_{\parallel}\int_{0}^{\infty}\mbox{d}c_{\perp}
c_{\perp}^{D-2}\Theta(b\leq 1)\Theta(c\leq
  1)\left. \MOPE{\GC}{\GA} \right|_{a=1,b,c}
\ .
\end{eqnarray}
We also have to derive an expression for the counter-term.
It can be written as
\begin{eqnarray}
  \label{4.20}
  I_{2}^{\ind{C}}(D,L)&=&-\frac{3}{2}\int_{a}\int_{c}a^{-\nu d}c^{-\nu d}\,\Theta(a\leq
  L)\Theta(c\leq L)
\ .
\end{eqnarray}
Using the same trick to derive w.r.t.\ the IR-regulator as
above, we obtain
\begin{eqnarray}
  \label{4.21}
2 \E  I_{2}^{\ind{C}}(D,L)&=&-\frac{3}{2}L^{2\E}\int_{\mathcal{A},\mathcal{B},\mathcal{C}}a^{-\nu d}c^{-\nu
  d}\,\Big[\delta(a-1)\Theta(c\leq a)+\delta(c-1)\Theta(a\leq c)\Big]\
  .
\end{eqnarray}\begin{figure}[t]
  \begin{center}
    \setlength{\unitlength}{1cm}
    \begin{picture}(7,5)
    \put(0,0){\raisebox{0mm}{\includegraphics[scale=0.5]{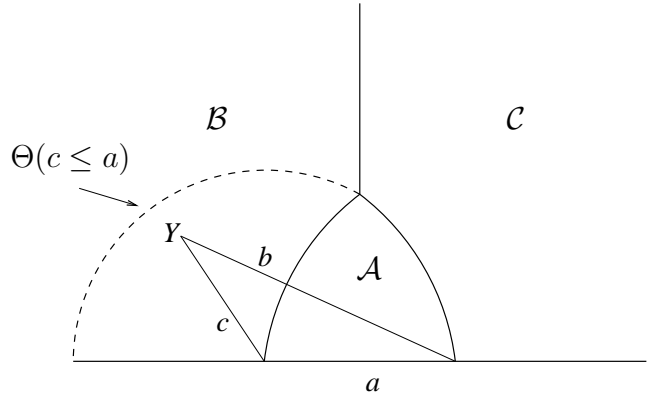}}} 
    \put(3.8,1.5){$\mathcal{A}$}
    \put(1.8,3.5){$\mathcal{B}$}
    \put(5.8,3.5){$\mathcal{C}$}
    \put(-0.8,3.0){$\Theta(c \leq a)$}
      \end{picture}
    \end{center} 
\caption{Area of integration (dashed) of the first integral in
    (\ref{4.21}). The intersection with $\mathcal{B}$ is to be mapped onto $\mathcal{A}$}
    \label{ftwol.0}
\end{figure}%
Using the technique of conformal mapping, which is presented in
appendix \ref{conformal},
both integrals above can be transformed into one having  support in the 
single sector $\mathcal{A}$ defined  through $b,c \leq a$, (where here $a$ is set
 to $1$). Analogously, we define sectors $\mathcal{B}$ and
$\mathcal{C}$, as the subset of domains where $b$ respectively $c$ is
the longest distance. Note that the first term in (\ref{4.21}) only has contributions
from $\mathcal{A}$ and $\mathcal{B}$,  while
the second can be restricted to  $\mathcal{A}$ and $\mathcal{C}$ after
renaming $a$ into $b$ and $c$ into $a$.\\
Performing explicitly the mapping of both $\mathcal{B}$ and $\mathcal{C}$
onto $\mathcal{A}$ one obtains 
\begin{eqnarray}
  \label{4.24}
  \int_{\mathcal{B}}(a^{-\nu d}c^{-\nu
  d}\Theta(c\leq a))\ts_{ a=1}&=&\int_{\mathcal{A}}b^{-\nu d-2\varepsilon}c^{-\nu
  d}\Theta(c\leq b)\nonumber\\
  \int_{\mathcal{C}}(a^{-\nu d}b^{-\nu
  d}\Theta(b\leq a))_{\mid a=1}&=&\int_{\mathcal{A}}c^{-\nu
  d-2\varepsilon}b^{-\nu d}\Theta(b\leq c) 
\ .
\end{eqnarray}
Combining the contributions from all sectors, we arrive at
\begin{equation}
  \label{4.25}
  2 \E I_2^{\ind{C}}(D,L)=-\frac{3}{2}L^{2\E}\int_{\mathcal{A}}\left(b^{-\nu d}{+}c^{-\nu
  d}{+}b^{-\nu d}c^{-\nu d}\mbox{max}(b,c)^{-2\varepsilon}\right)\ .
\end{equation}
Combining (\ref{4.18}) and (\ref{4.25}), we obtain an
expression for the complete diagram
\begin{eqnarray}
  \label{4.18.1}
  2\E   I_2(D,L)&=&3 L^{2\E}\frac{S_{D-1}}{S_{D}}\int_{-
  \infty}^{\infty}\mbox{d}c_{\parallel}\int_{0}^{\infty}\mbox{d}c_{\perp}
c_{\perp}^{D-2}\Theta(b\leq 1)\Theta(c\leq
  1)\nn\\
&& \times\Bigg\{
\left[\frac{1}{4}(1{+}b^{\nu}{+}c^{\nu})(b^{\nu}{+}c^{\nu}{-}1)(1{+}c^{\nu}{-}b^{\nu})(1{+}b^{\nu}{-}c^{\nu})\right]^{{-}d/2} \nn\\
&&   
\qquad -\half \left[b^{-\nu d}{+}c^{-\nu d}{+}b^{-\nu
d}c^{-\nu d}\mbox{max}(b,c)^{-2\varepsilon} \right] \Bigg\} \ .
\end{eqnarray}%
\begin{figure}[h,b]
\begin{center}
  \begin{minipage}[h]{4cm}
    \hspace{-20mm}
    {\raisebox{0mm}{\includegraphics[scale=0.4]{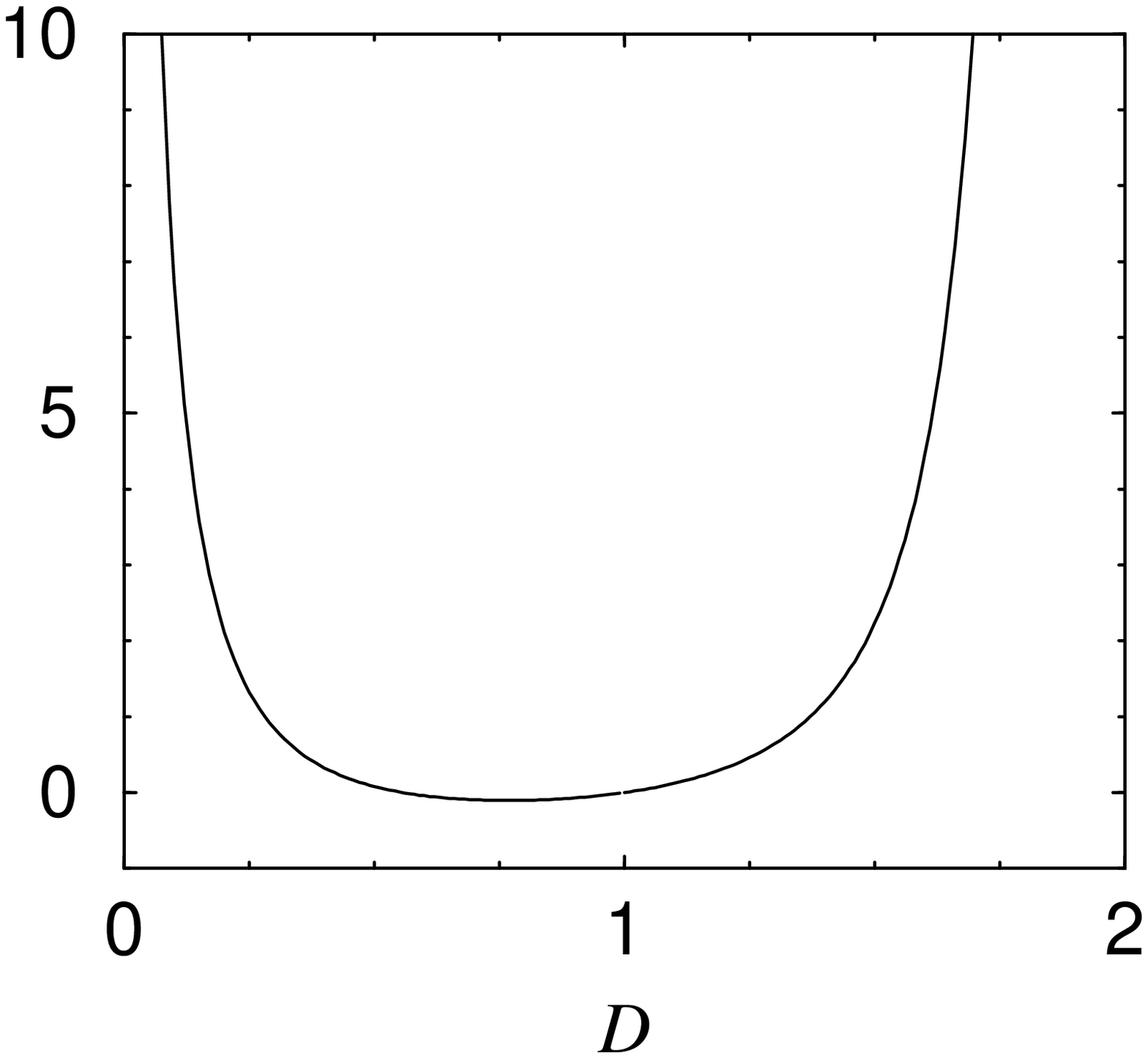}}}
  \end{minipage}
\begin{minipage}[h]{4cm}
\begin{tabbing}
  \hspace{40mm}
  \begin{tabular}[hc]{|c|c|}
    \hline
    D &  \\
    \hline
    0.1 & 6.71 \\
    0.2 & 2.11\\
    0.3 & 0.847\\
    0.4 & 0.325\\
    0.5 & 0.0807 \\
    0.6 & -0.0415\\
    0.7 & -0.0949 \\
    \hline
  \end{tabular} 
 
 \begin{tabular}[hc]{|c|c|}
    \hline
    D &  \\
    \hline
    0.8 & -0.102\\
    0.9 & -0.0712 \\
    1.0 & 0\\
    1.1 & 0.123\\
    1.2 & 0.321\\
    1.3 & 0.644\\
    1.4 & 1.20\\
    \hline
  \end{tabular} 
 
 \begin{tabular}[hc]{|c|c|}
    \hline
    D &  \\
    \hline
    1.5 &  2.24\\
    1.6 & 4.45\\
    1.7 & 10.4\\
    1.8 & 36.7\\
    1.9 & 658\\
    2.0 & $\infty$ \\
        &     \\
    \hline
  \end{tabular}
\end{tabbing}
\end{minipage}
\end{center}
     \caption{Numerically obtained results for the two-loop diagram in the MS-scheme. Some values are
    tabulated.}
    \label{f4.1}
\end{figure}
This expression will also be used later in a massive
scheme. Since here we are only interested in the $\E$-expansion, we
take the  limit of $\E\to 0$  in the integrand. We have made the
(non-trivial) check that
the integrand remains a finite integrable function, which can be 
integrated numerically:
\begin{eqnarray}
  \label{4.18.2}
   I_2(D,L)&=&3 \frac{L^{2\E}}{2\E} \frac{S_{D-1}}{S_{D}}\int_{-
  \infty}^{\infty}\mbox{d}c_{\parallel}\int_{0}^{\infty}\mbox{d}c_{\perp}
c_{\perp}^{D-2}\Theta(b\leq 1)\Theta(c\leq
  1)\nn\\
&& \times\Bigg\{
\left[\frac{1}{4}(1{+}b^{\nu}{+}c^{\nu})(b^{\nu}{+}c^{\nu}{-}1)(1{+}c^{\nu}{-}b^{\nu})(1{+}b^{\nu}{-}c^{\nu})\right]^{{-}d_{c}/2} \nn\\
&&   
\qquad -\half \left[b^{-\nu d_{c}}{+}c^{-\nu d_{c}}{+}b^{-\nu
d_{c}}c^{-\nu d_{c}} \right] \Bigg\} \nn\\
&& + O (\E^0)
\ .
\end{eqnarray}
Let us now proceed with the explicit numeric evaluation  of 
(\ref{4.18.2}). There are integrable singularities both in $c_{\parallel}$ and
$c_{\perp}$. To disentangle them \cite{WieseDavid1995,WieseDavid1997} we make a
change of variables from Cartesian coordinates $(c_{\parallel},c_{\perp})$ to 
radial and angular coordinates $(c,\alpha)$ according to 
\begin{equation}
  c_{\parallel} =c \cos(\alpha)  \ , \qquad 
  c_{\perp} = c \sin(\alpha)\ .
\end{equation}
such that
\begin{equation}
  \label{4.26}
  \int_{-\infty}^{\infty}\mbox{d}c_{\parallel}\int_{0}^{\infty}\mbox{d}c_{\perp}c_{\perp}^{D-2}\longrightarrow\
  \int_{0}^{1}\mbox{d}c\, c^{D-1}\int_{0}^{\pi/2}\mbox{d}\alpha\
 (\sin(\alpha))^{D-2}\ ,
\end{equation}
where the upper bounds of $\pi /2$ on $\A$ and $1$ on $c$ is due to the bound of $1$ on 
$b$ and $c$. We can further restrict integration to the half-sector
with $c<b$. In this sector, 
there remain singularities for small $\A$ and small $c$. 
The reader is invited to verify that they 
 are eliminated by a second change of variables
\begin{eqnarray*}
  \alpha&=&\frac{\pi}{2}\beta^{\frac{1}{D-1}}\\
  c&=&\tilde c^{\frac{1}{2-D}}\ .
\end{eqnarray*}
The above method works for $D>1$ only. For $D<1$, one has to
analytically continue the integration over $c_{\perp } $ in
(\ref{4.18}).  This can be done by partial integration, since the
integrand does not explicitly depend on $c_{\perp }$ but on
$c=\sqrt{c_{\perp }^{2}+c_{\parallel }^{2}}$. To have no boundary
terms, one should partially integrate before mapping everything onto
the sector $\cal A$.
 We have
explicitly checked consistency with the earlier used measure for $1{<}D{<}2$. Note finally,
that in the special case of $D=1$ in (\ref{4.18.2}) the integration measure
becomes a distribution reducing the integration measure to a line along
$c_{\parallel}$.\\

The diagram is shown on figure (\ref{f4.1}). It grows as $D^{-1}$ for
$D\to 0$ and  exponentially for $D\to 2$:
\begin{equation}
  \label{4.31}
  I_{2}(D)\sim \left(\frac{3}{4}\right)^{-d_{c}/2}\sim  \left(\frac{3}{4}\right)^{-D/(2{-}D)}\ .
\end{equation}
In the case of polymers $D{=}1$ it vanishes. The reason is that 
 the diagrams factorize, and thus the 1-loop result
is correct to all orders when working in an appropriate scheme. (See
e.g. \cite{WieseHabil}.)

\subsection{RG-function and extrapolation}%
\begin{figure}[t]
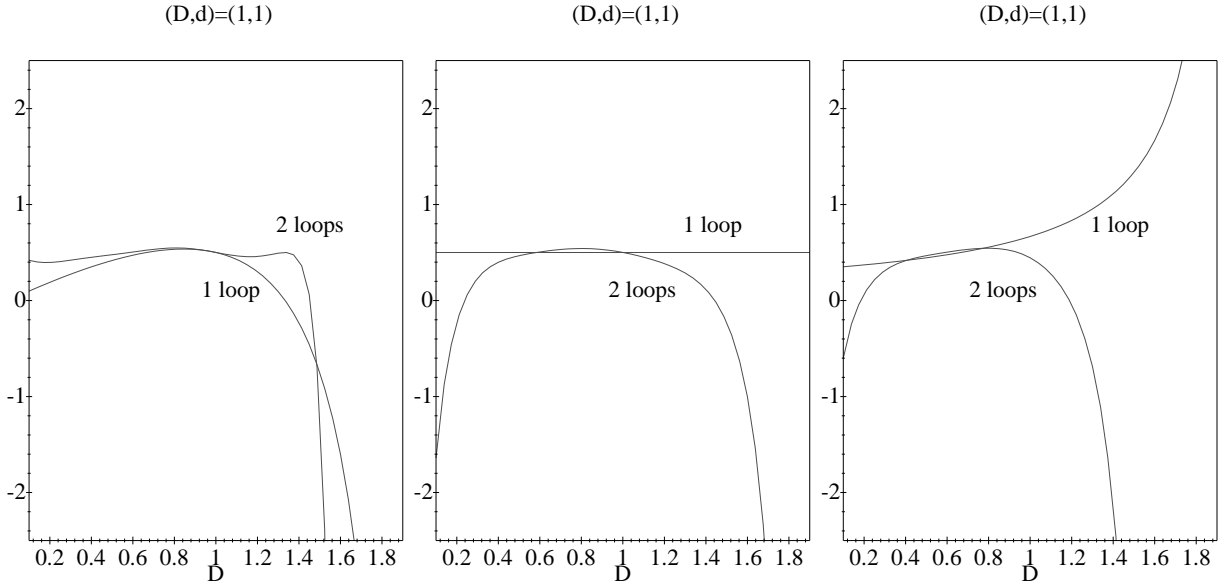

  \begin{center}
    \hspace{-2.2cm}
    \begin{minipage}[h]{3.2cm}
      \raisebox{0mm}{\includegraphics[scale=0.35]{./eps/o11dD.epsi}} 
    \end{minipage}
    \hspace{2cm}
    \begin{minipage}[h]{3.2cm}
      \raisebox{0mm}{\includegraphics[scale=0.35]{./eps/o11Deps.epsi}} 
    \end{minipage}
    \hspace{2cm}
\begin{minipage}[h]{3.2cm}
      \raisebox{0mm}{\includegraphics[scale=0.35]{./eps/o11DDc.epsi}} 
    \end{minipage}
  \end{center}
  \begin{center}
  \caption{Correction to scaling exponent $\omega$ for $(D{=}1,d{=}1)$
with different extrapolation parameters $(D,d),
(D,\varepsilon(D,d)),(D,D_{c}(d))$ (from left to right). The abscissa
labels the corresponding internal dimension $D_{0}$ of the departure
point on the critical line $\varepsilon(D_{0},d_{0}){=}0$.}
  \label{f4.2}
  \end{center}
\end{figure}%
\begin{figure}[htbp]
  \begin{center}
    \hspace{-2.2cm}
    \begin{minipage}[h]{3.2cm}
      \raisebox{0mm}{\includegraphics[scale=0.35]{./eps/o21DDc.epsi}} 
    \end{minipage}
    \hspace{2cm}
    \begin{minipage}[h]{3.2cm}
      \raisebox{0mm}{\includegraphics[scale=0.35]{./eps/o22DDc.epsi}} 
    \end{minipage}
    \hspace{2cm}
    \begin{minipage}[h]{3.2cm}
      \raisebox{0mm}{\includegraphics[scale=0.35]{./eps/o23DDc.epsi}} 
    \end{minipage}

    \hspace{-2.2cm}
     \begin{minipage}[h]{3.2cm}
      \raisebox{0mm}{\includegraphics[scale=0.35]{./eps/o21dD.epsi}} 
    \end{minipage}
    \hspace{2cm}
    \begin{minipage}[h]{3.2cm}
      \raisebox{0mm}{\includegraphics[scale=0.35]{./eps/o22dD.epsi}} 
    \end{minipage}
    \hspace{2cm}
    \begin{minipage}[h]{3.2cm}
      \raisebox{0mm}{\includegraphics[scale=0.35]{./eps/o23dD.epsi}} 
    \end{minipage}
    \hspace{-3cm}
    \begin{minipage}[h]{2cm}
      \begin{tabbing}
        \begin{tabular}[h]{|c|c|c|c|c|}
          \hline
          $(x,y)\ ; (D,d)$ & (1,1) & (2,1) & (2,2) & (2,3)  \\
          \hline
          $(D,Dc)\ ,\ \mbox{max}$ & 0.55 & 2.80 & 2.43 & 2.10  \\
          \hline
          $(D,d)\ ,\ \mbox{max}$ & 0.55 & 2.80 & 2.43 & 2.20  \\
          \hline
          $(D,\varepsilon)\ ,\ \mbox{max}$ & 0.55 & & &  \\
          \hline
        \end{tabular} 
      \end{tabbing}
    \end{minipage}

 \end{center}

  \begin{center}
  \caption{Correction to scaling exponent $\omega$ for different
extrapolation points $(D,d)$: Extrapolation parameters are
$(D,D_c(d))$ (top) and $(D,d)$ (bottom). The abscissa labels the
corresponding internal dimension $D_{0}$ of the departure point on the
critical line $\varepsilon(D_{0},d_{0}){=}0$. Furthermore, the
two-loop values with minimal sensitivity with respect to the departure
point are tabulated.}  \label{f4.21}
  \end{center}
\end{figure}

>From (\ref{3.22}) the renormalization $\beta$-function in second order of
perturbation theory reads
\begin{equation}
  \label{4.32}
  \beta(g)=-\varepsilon g+\frac{g^{2}}{2}-\frac{g^{3}}{3}I_{2}(D)+O(g^{4})\ .
\end{equation}
The long-distance behavior of the theory is governed by the IR--stable
fixed point of the RG--flow. Fixed points are given by zeros of the
$\beta$--function:
\begin{equation}
  \label{4.33}
  \beta(g^{*})=0\qquad \Rightarrow \quad g^{*}=0\quad \vee \quad
  -\varepsilon+\frac{g^{*}}{2}-\frac{g^{*2}}{3}I_{2}(D)+O(g^{*3})=0\ .
\end{equation}
The physically interesting nontrivial fixed point is the one that is 
next to zero. In an $\varepsilon$--expansion it reads:
\begin{equation}
  \label{4.35}
  g^{*}=2\varepsilon+\frac{8}{3}I_{2}(D)\varepsilon^{2}+O(\varepsilon^{3})\ .
\end{equation}
The correction to scaling exponent $\omega$ at this fixed point is found in an
$\varepsilon$-expansion to be
\begin{equation}
  \label{4.36}
  \omega(g^{*})=\varepsilon-\frac{4 }{3}I_{2}(D)\varepsilon^{2}+O(\varepsilon^{3})\ .
\end{equation}
The question now arises of how to calculate $\omega $ for a given physical
situation. 
The general strategy that we follow in the MS-scheme is that we make
use of the freedom to choose any point on the critical line
(\ref{f2.1}) around which to start our expansion
\cite{Hwa1990,WieseDavid1995}.  To expand towards some physically
interesting point $(D,d)$ we furthermore dispose of the freedom to
follow any path starting on the curve $\varepsilon{=}0$. Since our
expansion (\ref{4.36}) is exact in $D$ but only of second order in
$\E$, we can equally well expand it to second order in $D$ around any
point on the critical curve.  Now we can change our extrapolation path
through an invertible transformation $(x,y)=(x(D,d),y(D,d))$. One
expresses $D$ and $\varepsilon$ as functions of $x$ and $y$ and
reexpands $\omega$ to second order in $x$ and $y$ around the point
$(x_{0},y_{0})=(x_{0}(D_{0},d_{c}(D_{0})),y_{0}(D_{0},d_{c}(D_{0})))$,
where we recall that $d_c (D)$ is defined such that $\E(D,d_c(D))=0$.
The aim is to find an optimal choice of variables $(x,y)$. The
guidelines for such a choice have been discussed in
\cite{WieseDavid1997}, where this procedure has been successfully
applied to extrapolate results for the anomalous dimension of
self-avoiding membranes.  As a general rule, we ``trust'' most a
result that is the least sensitive to the starting point of the
extrapolation. Therefore, we would like to find a plateau for some
suitable chosen variables $(x,y)$. This procedure works well for
polymers, or more generally for membranes with inner dimension close
to 1.  However it turns out that for membranes it works much less well
than in the self-avoiding case: we could not find a plateau, but at
best an extremum.

Some extrapolations are shown in the following figures (\ref{f4.2},
\ref{f4.21}). We start with polymers
interacting with a $\delta$-wall corresponding to the point $(D,d)=(1,1)$.
(Note that for $d<2$ the interaction becomes relevant for polymers.) This can
be seen on fig.\ \ref{f4.2}. The values obtained from the plateaus appearing
at the two-loop level $(\omega{\approx} 0.55)$ are quite close to the exact
value $\omega=1/2$, known from the fact that the result in $D=1$ at
one-loop is exact. From the latter it follows that it is best to use
$(D,\varepsilon)$ as extrapolation parameters and to stay in $D=1$.

Let us now turn to the results obtained for $\omega$ extrapolated to points
$D=2$ (fig.\ \ref{4.21}): The extrapolated value depends strongly on
the starting-point on the critical line. We can identify a least sensitive value,
which lies below $D=1$, reflecting the $D$-dependence of the two-loop
diagram. We obtain identical results for different extrapolation
paths. Note, that if we fix $d=1$ and perform an
$\varepsilon$-expansion in $D$ as is commonly done \cite{LassigLipowsky1993}, then, this
corresponds to starting from $D_{c}(d{=}1)=2/3<1$. The $d$-dependence of
the extrapolated values is recovered in the massive scheme, which we discuss
in the next section.


\section{Calculation in fixed dimension: The massive scheme}
\label{massive}
\subsection{Phantom manifolds}
\begin{figure}[hhhh]
\centerline{\hfill \parbox{0.6\textwidth}
        {\fig{0.6\textwidth}{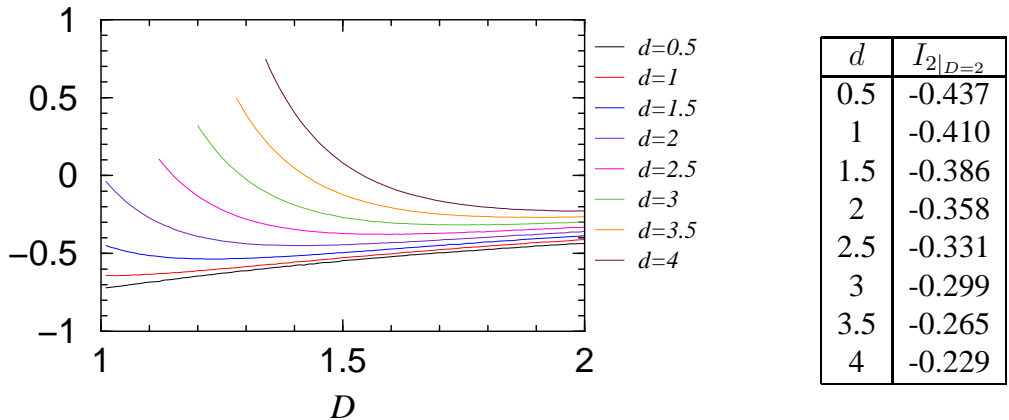}}
\hfill 
\begin{tabular}[hc]{|c|c|}
          \hline
          $d$ & $I_{2|_{D{=}2}}$ \\
          \hline
          0.5 & -0.437 \\
          1 & -0.410\\
          1.5 & -0.386\\
          2 & -0.358\\
          2.5 & -0.331 \\
          3 & -0.299\\
          3.5 & -0.265 \\
          4 &   -0.229 \\
          \hline
\end{tabular} \hfill 
}
\caption{2-loop diagram in fixed dimension for embedding dimensions
       $d{=}0.5$ to $d{=}4$ (from bottom to top). The curves are only shown
       for values of $D$ such that $\E{\geq}0$.}
    \label{f5.1}
\end{figure}

An alternative scheme to an $\varepsilon$-expansion is a calculation
in fixed dimension. This way we hope to circumvent the difficulty that
we faced when calculating the two-loop diagram in the previous section
in the limit of $D\to 2$: Since the diagram behaves like
$\left(\frac{4}{3} \right)^{d/2}$, it becomes large when the dimension
of the embedding space is large. When working in an $\E$-expansion,
$d$ and $D$ are related by $\E=D-d(2-D)/2$ and sending $\E\to0 $
before $D\to2$ imposes that $d\to \infty$. Thus one would like to work
at finite $d$, thus finite $\E$.  The diagrams are UV-convergent for
$d<d_{c}$. They are cut-off in the infrared by a finite membrane
size. Alternatively, we can use a ``soft'' IR cut-off, that is we
introduce a chemical potential $\tau$ into the Hamiltonian and sum over
all manifold sizes. For polymers $(D=1)$ this scheme corresponds to a
``massive scheme'' in an $O(N)$-symmetric, scalar $\phi^{4}$-theory as
can be seen via a de Gennes Laplace transformation on the $N=0$
component limit. (The correlator in momentum-space reads
$\frac{1}{k^{2}+t}$) In the following we keep a hard IR regulator and
calculate the two-loop diagram in fixed dimension. We use the
analytical expression given in (\ref{4.18.1}).  For the numerical
integration in $1<D<2$ we proceed as before. The result for different
embedding dimensions $d$ is shown graphically on figure \ref{f5.1},
where also some explicit values in the limit of $D=2$ are given. Note,
that the exact factorization-property in the limit of $D=1$ is only valid in
the MS-scheme or in a massive scheme with suitable soft cut-off as
mentioned above. However, even the method is not tailored to capture
that feature, the obtained value $\omega=0.6$ in the case of
$(D,d)=(1,1)$ is not too far from the exact value $1/2$. We expect
that higher order terms would again reproduce the exact value of
$\omega =1/2$. Also note that results for
membranes ($D=2$) are close to those  from the $\E$-expansion,
especially for small $d$.
\begin{figure}[htbp]
\centerline{\hfill 
\parbox{0.6\textwidth}{\fig{0.6\textwidth}{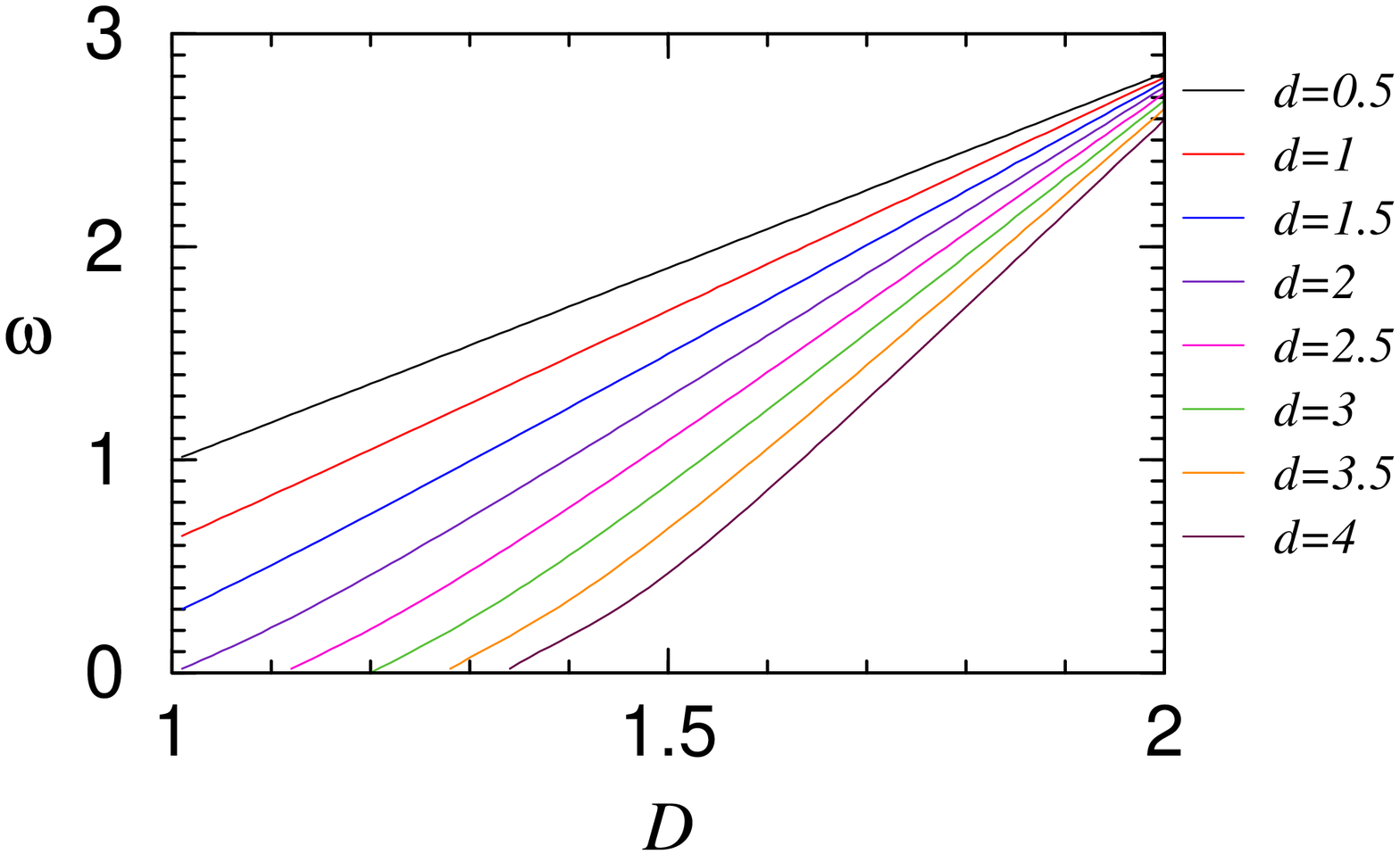}}
\hfill 
\begin{tabular}[hc]{|c|c|}
          \hline
          $d$ & $\omega_{|_{D{=}2}}$ \\
          \hline
          0.5 & 2.82 \\
          1 & 2.80\\
          1.5 & 2.77\\
          2 & 2.75\\
          2.5 & 2.72 \\
          3 & 2.69\\
          3.5 & 2.65 \\
          4 &   2.60 \\
          \hline
\end{tabular} \hfill }
\caption{Correction to scaling exponent $\omega$  for embedding
dimension $d{=}0.5$ to $d{=}4$ 
    (from top to bottom). $\E$ is restricted to $\E{\geq}0$; othervise
one has to work with an explicit UV-cutoff.} 
  \label{f5.2}
\end{figure}

\subsection{Self-avoiding membranes}
Up to now we were considering a $D$-dimensional polymerized
``phantom'' membrane interacting with a point-like impurity in
$d$-dimensional embedding space modeled through a
$\delta$-potential. The infinite fractal dimension of the free manifold was
causing  problems, and since all operators of the type
$(-\Delta_{r})^{n}\delta(r),\ n{\geq}0$, attain the same relevance in
$D=2$, it is not clear whether the model will break down. Physically,
the problem is much better defined for self-avoiding membranes. From
field-theoretical calculations, which have recently been refined up to
 two-loop  \cite{DavidWiese1996,WieseDavid1997}, we know that two-dimensional
self-avoiding   manifolds embedded in three dimensional
space have an anomalous dimension of $\nu^{*}\approx 0.85$.  We now
try to overcome the problem of an infinite Hausdorff-dimension by a
crude simplification: We approximate the self-avoiding manifold by a
Gaussian theory with the same scaling dimension, that is a Hamiltonian
of the type \cite{DDG1}
\begin{equation}
  \label{5.2}
  \mathcal{H}[r] = \frac{1}{k{-}D}\int_{x}\
  \frac{1}{2}\,r(x)(-\Delta_{x})^{k/2}r(x)\ ,
\end{equation}
where
\begin{displaymath}
  r:\ x \in \R^{D} \longrightarrow  r( x) \in \R^{d}\ 
\end{displaymath}
as before providing the two-point correlator in our normalizations:
\begin{equation}
  \label{5.3}
  C(x,y)= |x{-}y|^{k-D}\ .
\end{equation}
The scaling dimension is
\begin{equation}
  \label{5.4}
  \nu=\frac{k{-}D}{2}
\ .
\end{equation}
We recover (\ref{3.7}) by setting $k=2$, but the model can be 
continued analytically to any real value of $k$, with $k\ge2$.
Setting
\begin{equation}
  \label{5.5}
  k^{*}=2\nu^{*}+D
\end{equation}
we get a Gaussian manifold with identical scaling dimension as obtained for
self-avoiding crumpled membranes at two-loop. The critical embedding dimension
of the interaction then reads
\begin{equation}
  \label{5.6}
  d_{c}=\frac{2D}{k^{*}{-}D}=\frac{D}{\nu^{*}}\ ,
\end{equation}
that is the interacting is relevant in $d{<}d_{c}$. For membranes
$d_{c}{=}\frac{2}{0.85}\approx 2.4$.
All operators of the type
$(-\Delta_{r}^{n}\delta^{d}(r))$ are at least naively irrelevant for $n>0$, as 
\begin{equation}
  \label{5.7}
  [(-\Delta_{r}^{n}\delta^{d}(r))]=-\nu^{*} (d+2n)\ 
\end{equation}
and the corresponding coupling has in inverse length units the dimension
\begin{equation}
  \label{5.8}
  [g^{(n)}_{0}]=D-\nu^{*} (d+2n)<0\ ,\qquad n>0\ .
\end{equation}
We may repeat the calculation in the massive scheme setting the free scaling
dimension $\nu^{*}{=}0.85$ and $(D,d){=}(2,1)$. The result is shown in table
(\ref{t5.1}).\\
Let us stress that this method can not be turned into a systematic expansion,
since perturbation theory neglects all effects of the non-Gaussian nature of
the nontrivial crumpled state. We only know that for $d\to \infty$ the
Gaussian variational approximation for the Hamiltonian, i.e.\ (\ref{5.2})
becomes exact \cite{Goulian1991,LeDoussal1992,GuitterPalmeri1992}. 
\begin{table}[htbp]
  \begin{center}
    \begin{minipage}[h]{3.2cm}
      \begin{tabbing}
        \hspace{2cm}
        \begin{tabular}[h]{|c|c|c|c|}
          \hline
                & I & $\omega(g^{*})$ & $\theta(g^{*})$   \\
          \hline
          $1$-loop & 1.000 & 1.15 & 2.71  \\
          \hline
          $2$-loop & -0.379  & 1.49 & 3.10  \\
          \hline
        \end{tabular} 
      \end{tabbing}
    \end{minipage}
    \label{t5.1}
  \end{center}
  \caption{Two-loop results for self-avoiding membranes $(D{=}2)$ approximated through a
                Gaussian theory having identical scaling dimension
                $\nu^{*}{=}0.85$. Results are shown for the corresponding
                diagram, the correction to scaling exponent $\omega$ and the
                contact exponent $\theta$.}
\end{table}


\section{Summation of the perturbation series in the limit $D=2$}
In the last section we introduced a ``massive'' scheme allowing to
perform the renormalization procedure in fixed dimension. As can be
seen in figures \ref{f5.1} and \ref{f5.2}, the limit $D\to 2$ can be
taken in the 2-loop diagram and is smooth.  Surprisingly, this limit
can be calculated analytically: Setting
$\nu=0$ in  (\ref{4.18.1}), the integrand becomes constant, and thus
trivial to calculate. This is a
striking property. Even more, if we slightly modify the regularization
prescription, this property holds to {\em all loop orders} in
perturbation theory. Below, we will give an analytic
expression for the complete perturbation series, which allows for
analyzing the strong coupling limit. Furthermore, we show how
corrections in $(2-D)$ can be incorporated. We will see that only the
latter depend on the explicit cut-off procedure, a point which will be
further clarified below. As a non-trivial check, when extrapolating back to
polymers ($D=1$), the corresponding result for $\omega $ is
approximately reproduced. 

\subsection{$N$-loop order}
In order to calculate the $N$--loop contribution to the perturbation
theory we need 
the SDE of $N+1$ contact interactions. The normal ordered product of the
corresponding vertex operators reads
\begin{eqnarray}
  \label{6.1}
  \rme^{ik_{1}r(x_{1})}\rme^{ik_{2}r(x_{2})}\cdot \dots \cdot
  \rme^{ik_{N+1}r(x_{N+1})}\ =\ \bn
\rme^{i(\sum_{n=1}^{N+1}k_{n}r(x_{n}))}\en \ 
  \prod_{i,j=1}^{N+1}\rme^{\frac{1}{2}k_{i}k_{j}C(x_{i}{-}x_{j})}\ .
\end{eqnarray}
We choose $x_{N{+}1}$ as  root in the renormalization procedure (for the
precise definition, see below) and therefore $k_{N{+}1}$
to integrate 
over in order to obtain the $\delta $-distribution (or its
derivative), onto   which to
project. We thus shift 
\begin{displaymath}
  k_{N{+}1}\ \longrightarrow \ k_{N{+}1}-\sum_{j=1}^{N}k_{j}
\end{displaymath}
and rewrite the quadratic form in (\ref{6.1}) as
\begin{eqnarray}
  \frac{1}{2}\sum_{i,j=1}^{N+1}&&k_{i}k_{j}C(x_{i}{-}x_{j})=\sum_{j=1}^{N}k_{N{+}1}k_{j}C(x_{N{+}1}{-}x_{j})+\frac{1}{2}\sum_{i,j=1}^{N}k_{i}k_{j}C(x_{i}{-}x_{j})\nn\\
  &&\longrightarrow \
  \sum_{j=1}^{N}k_{N{+}1}k_{j}C(x_{N{+}1}{-}x_{j})-\sum_{i,j=1}^{N}k_{i}k_{j}C(x_{N{+}1}{-}x_{j})+\frac{1}{2}\sum_{i,j=1}^{N}k_{i}k_{j}C(x_{i}{-}x_{j})\nn \\
  &&=\sum_{j=1}^{N}k_{N{+}1}k_{j}C(x_{N{+}1}{-}x_{j})-\sum_{i,j=1}^{N}k_{i}k_{j}\frac{C(x_{N{+}1}{-}x_{i})+C(x_{N{+}1}{-}x_{j})-C(x_{i}{-}x_{j})}{2}\ .\nn\\
\label{6.2.a}
\end{eqnarray}
(\ref{6.1}) then becomes (up to subdominant terms\footnote{Here we
argue in terms of an OPE. Note 
that in order to make these arguments rigorous, one should ask of
whether these terms affect the effective action at {\em  constant}
background-field. This is not the case, since then $\nabla r=0$.} involving
spatial derivatives of $r$)
\begin{eqnarray}
&&\bn \exp\left[ik_{N+1}r (x_{N+1}) \right]\en \nn \\
&&\quad\times\exp\left[\sum_{j=1}^{N}k_{N{+}1}k_{j}C(x_{N{+}1}{-}x_{j})-\sum_{i,j=1}^{N}k_{i}k_{j}\frac{C(x_{N{+}1}{-}x_{i})+C(x_{N{+}1}{-}x_{j})-C(x_{i}{-}x_{j})}{2}  \right]\nn\\
\end{eqnarray}
and the integration over $k_{N+1}$ produces the $\delta$-interaction
and its derivatives; the latter  come from the first term in
(\ref{6.2.a}). They are subdominant and thus will be dropped. We abbreviate the
quadratic form in (\ref{6.2.a}) as $(D_{ij})$, where the matrix elements are
\begin{eqnarray}
  \label{6.2}
  D_{ij} &=&
  \frac{1}{2}\left[C(x_{N{+}1}{-}x_{i})+C(x_{N{+}1}{-}x_{j})-C(x_{i}{-}x_{j})\right]\ ,\quad
  i\not=j\ ,\nonumber\\
  D_{ii} &=& C(x_{N{+}1}{-}x_{i})\ .
\end{eqnarray}
Integrating out the momenta $k_{1},\dots k_{N}$ yields the
operator product expansion of $N+1$ contact interactions as
\begin{eqnarray}
  \label{6.3}
  \tilde \delta^{d}(r(x_{1}))\, \tilde \delta^{d}(r(x_{2}))\,\dots\,\tilde \delta^{d}(r(x_{N+1})) &=&
  [\det (D_{ij})]^{-d/2}\, \tilde \delta^{d}(r(x_{N{+}1}))\nonumber \\
&& +\,\textrm{subdominant operators}\ .
\end{eqnarray}

\subsection{The limit $D\to 2$ and $(2{-}D)$-expansion}
\begin{figure}[t]
\begin{center}
\raisebox{-10mm}{\includegraphics[scale=0.3]{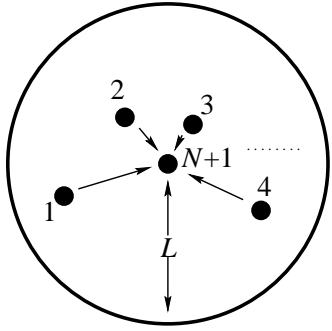}}
\end{center}
\caption{Regularization scheme for the $N$-loop diagrams}
    \label{f6.1}
\end{figure}

Whereas (\ref{6.3}) is still completely general, we now want to study
the limit $D\to2$ as well as perturbations above it.  We start by
remarking that the matrix $(D_{ij})$ may be expanded in powers of
$2{-}D$:
\begin{equation}
  \label{6.4}
  D_{ij}=\sum_{n=0}^{\infty}\frac{(2{-}D)^{n}}{n!}D_{ij}^{(n)}\ ,
\end{equation}
where
\begin{eqnarray}
  \label{6.5}
  D_{ij}^{(n)}&=&\frac{\mu^{-2\nu}}{2}(\ln^{n}|(x_{N{+}1}{-}x_{i})\mu|+\ln^{n}|(x_{N{+}1}{-}x_{j})\mu|-\ln^{n}|(x_{i}{-}x_{j})\mu|)\
  ,\quad i\not= j\nonumber\\
  D_{ii}^{(n)}&=&\mu^{-2\nu}\ln^{n}|(x_{N{+}1}{-}x_{i})\mu|\ ,
\end{eqnarray}
$\mu=1/L$.
Then,
\begin{eqnarray}
  \label{6.6}
  \det (D)&&=\exp[\mbox{Tr}\ln (D_{ij})]\nonumber\\
  &&=\det(D^{(0)})\cdot\exp[\mbox{Tr}\ln[\mathbf{1}+\sum_{n=1}^{\infty}\frac{(2{-}D)^{n}}{n!}[D^{(0)}]^{{-1}}D^{(n)}]]\ ,
\end{eqnarray}
where $[D^{(0)}]^{{-1}}$ denotes the inverse matrix to $D^{(0)}$.
Let us expand the OPE-coefficient in (\ref{6.3}) up to second order in
$2-D$. Expanding first the logarithm,
\begin{eqnarray*}
  \lefteqn{\ln[\mathbf{1}+\sum_{n=1}^{\infty}\frac{(2{-}D)^{n}}{n!}[D^{(0)}]^{-1}D^{(n)}]}\\
  &&=(2{-}D)[D^{(0)}]^{{-1}}
 D^{(1)}+\frac{(2{-}D)^{2}}{2}\left[[D^{(0)}]^{{-1}} 
D^{(2)}-([D^{(0)}]^{{-1}} D^{(1)})^{2}\right]+O((2{-}D)^{3})\ ,
\end{eqnarray*}
and inserting into (\ref{6.6}) one arrives at
\begin{eqnarray}
  \label{6.7}
  &&[\det(D)]^{-d/2}=[\det(D^{(0)})]^{-d/2}\left[1-\frac{(2{-}D)d}2\,
  \mbox{Tr}[D^{(0)}]^{{-1}} D^{(1)}\right.\nonumber\\
  &&\left.-\frac{(2{-}D)^{2}d}4\left[\mbox{Tr}[D^{(0)}]^{{-1}}D^{(2)} -\mbox{Tr}([D^{(0)}]^{{-1}}D^{(1)})^{2}-\frac{d}{2}\mbox{Tr}^{2}[D^{(0)}]^{{-1}} D^{(1)}\right]+O((2{-}D)^{3})\right]\
  .\nonumber\\
\end{eqnarray}
The zeroth order is $[\det(D^{(0)})]^{-d/2}$. In order to proceed we
need the latter and the inverse matrix $[D^{(0)}]^{-1}$. Let us first
calculate the determinant. From (\ref{6.4}) we have
\begin{eqnarray}
  \label{6.8}
  D^{(0)}&=&\mu^{-2\nu}\left(
    \begin{array}{cccc}
      1 & \frac{1}{2} & \ldots & \frac{1}{2} \\
      \frac{1}{2} & 1 & \ddots & \vdots \\
      \vdots & \ddots & \ddots & \frac{1}{2} \\
      \frac{1}{2} & \ldots & \frac{1}{2} & 1
    \end{array}
\right)=\frac{\mu^{-2\nu}}{2}\left(\left(
    \begin{array}{ccc}
      1 & \ldots & 0 \\
      \vdots & \ddots & \vdots \\
      0 & \vdots & 1
    \end{array}
  \right)+\left(
    \begin{array}{ccc}
      1 & \ldots & 1 \\
      \vdots & \ddots & \vdots \\
      1 & \ldots & 1
    \end{array}
\right)\right) =\frac{\mu^{-2\nu}}{2}(\mathbf{1}+N\mathbf{P})\ ,\nonumber\\
   \mathbf{P}&=&\frac{1}{N}\left(\begin{array}{ccc}
    1 & \ldots & 1 \\
    \vdots & \ddots & \vdots \\
    1 & \ldots & 1
  \end{array}\right)
\end{eqnarray}
being the projector onto $(1,1,\dots,1)$. Denoting by
    $\mbox{dim Im(\bf{P})}$ the dimension of the space onto which $\mathbf{P}$
    projects one has the general formula:
\begin{displaymath}
  \det (\mathbf{1}+a\mathbf{P})=(1+a)^{\mbox{dim Im(\bf{P})}}\ . 
\end{displaymath}
Since here $\mbox{dim Im(\bf{P})}=1$, we find
\begin{equation}
  \label{6.9}
  \det(D^{(0)})=\mu^{-2\nu N}\,\frac{1+N}{2^{N}}\ .
\end{equation}
The inverse matrix of $(\mathbf{1}+N\mathbf{P})$ can be inferred from the ansatz
\begin{displaymath}
  (\mathbf{1}+N\mathbf{P})(\mathbf{1}+b\mathbf{P})=\mathbf{1}\ .
\end{displaymath}
Since $\mathbf{P}$ is a projector $\mathbf{P}^{2}=\mathbf{P}$ and this implies
\begin{displaymath}
  (N+b+Nb)\ \mathbf{P}=\mathbf{0}\ .
\end{displaymath}
Therefore, $b=-N/(N+1)$ and
\begin{equation}
  \label{6.10}
  [D^{(0)}]^{-1}=2\left(\mathbf{1}-\frac{N}{N+1}\mathbf{P}\right)\mu^{2\nu}\ .
\end{equation}
To obtain the first order contribution in $(2{-}D)$ from eq.\ (\ref{6.6}) we only need
$ [D^{(0)}]^{-1} D^{(1)}$, which reads
\begin{eqnarray}
  \label{6.11}
 \left( [D^{(0)}]^{-1} D^{(1)}\right)_{ij}=\left(2D_{ij}^{(1)}
-\frac{2}{1+N}\sum_{k=1}^{N}D_{ik}^{(1)}\right)\mu^{2\nu}\ . 
\end{eqnarray}
The trace of (\ref{6.11}) can easily be performed, with the result
\begin{eqnarray}
  \label{6.12}
  \mbox{Tr} (
[D^{(0)}]^{-1}D^{(1)})=\left(\frac{2N}{1+N}\sum_{i=1}^{N}D_{ii}^{(1)}-
\frac{2}{1+N}\sum_{i=1}^{N}\sum_{k=1}^{N}(1-\delta_{ik})D_{ik}^{(1)}\right)
\mu^{2\nu} 
\ .
\end{eqnarray}
In order to factorize the integrals as much as possible, we found it
convenient to modify the regularization prescription  (\ref{3.16}).
In  each order of perturbation theory we have to integrate the
expression  (\ref{6.3}) over internal distances. These integrals
have to be regularized in the infrared through an appropriate IR cut-off. In
the preceding sections we have cut off the integrals such that all relative
distances between internal points were smaller than  $L=1/\mu$. In
the following we replace this by the demand that all distances from
the fixed, but arbitrarily 
chosen point $x_{N{+}1}$, the root of the subtraction prescription,
are bounded by $L$. Other distances are 
not restricted (see fig.\ \ref{f6.1}). Alternatively, one could
calculate on a closed manifold of finite size. Of course, then the
correlator has to be modified. Work is in progress to implement this
procedure, which has the advantage of being more systematic
\cite{PinnowWieseProgress}. 

To simplify the calculations, we further introduce the following notation:
\begin{equation}
  \label{6.13.1}
  \overline{f(x_{i_{1}},\dots,x_{i_{k}})} := \int_{x_{1}}\cdots\int_{x_{N}}f(x_{i_{1}},\dots,x_{i_{k}})
\end{equation}
with new normalization
\begin{equation}
  \label{6.13.2}
  \int_{x}\ :=\ \frac{D}{S_{D}}\int\mbox{d}^{D}x\, \theta(|x|<1)\ ,
\end{equation}
which is chosen such that the overbar can be thought of as an
averaging procedure, and especially
\begin{equation}
  \label{6.13.3}
  \overline{1}=1\ .
\end{equation}
Thanks to our regularization prescription the integral of (\ref{6.12})
over internal points can be replaced by $\mu^{-ND}$ (for the
integration measure $L^{ND}$) times 
\begin{eqnarray}
  \label{6.13}
  \overline{\mbox{Tr} ( [D^{(0)}]^{-1}D^{(1)})}&=&\frac{2N^{2}}{1+N}\ \overline{\ln|(x_{N{+}1}-x_{i})|}\nonumber\\
  &&-\left(\frac{2N(N-1)}{1+N}\right)\left(\overline{\ln|x_{N{+}1}-x_{i}|}-\frac{1}{2}\
  \overline{\ln|x_{i}-x_{j}|}\right)\nonumber\\
  &=&\frac{2N}{1+N}\ \overline{\ln|x_{N{+}1}-x_{i}|}+\frac{N(N-1)}{1+N}\ \overline{\ln|x_{i}-x_{j}|}\ ,
\ .
\end{eqnarray}
Introducing a diagrammatic notation
\begin{equation}
  \label{6.14}
  \raisebox{-4.5mm}{\includegraphics[scale=0.1]{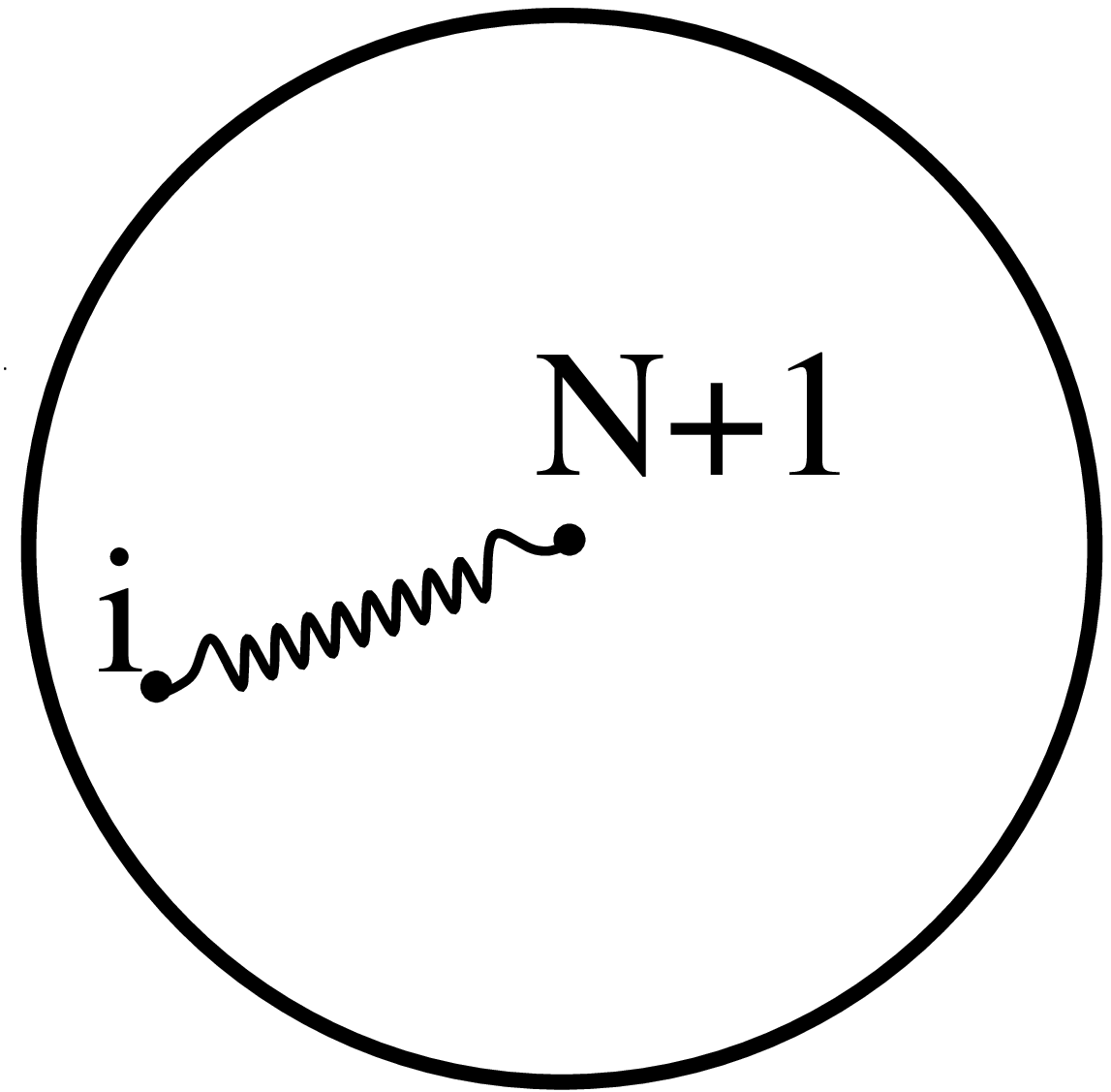}} :=
\overline{\ln |x_{N{+}1}{-}x_{i}|} 
\end{equation}
and
\begin{equation}
  \label{6.15}
  \raisebox{-4.5mm}{\includegraphics[scale=0.1]{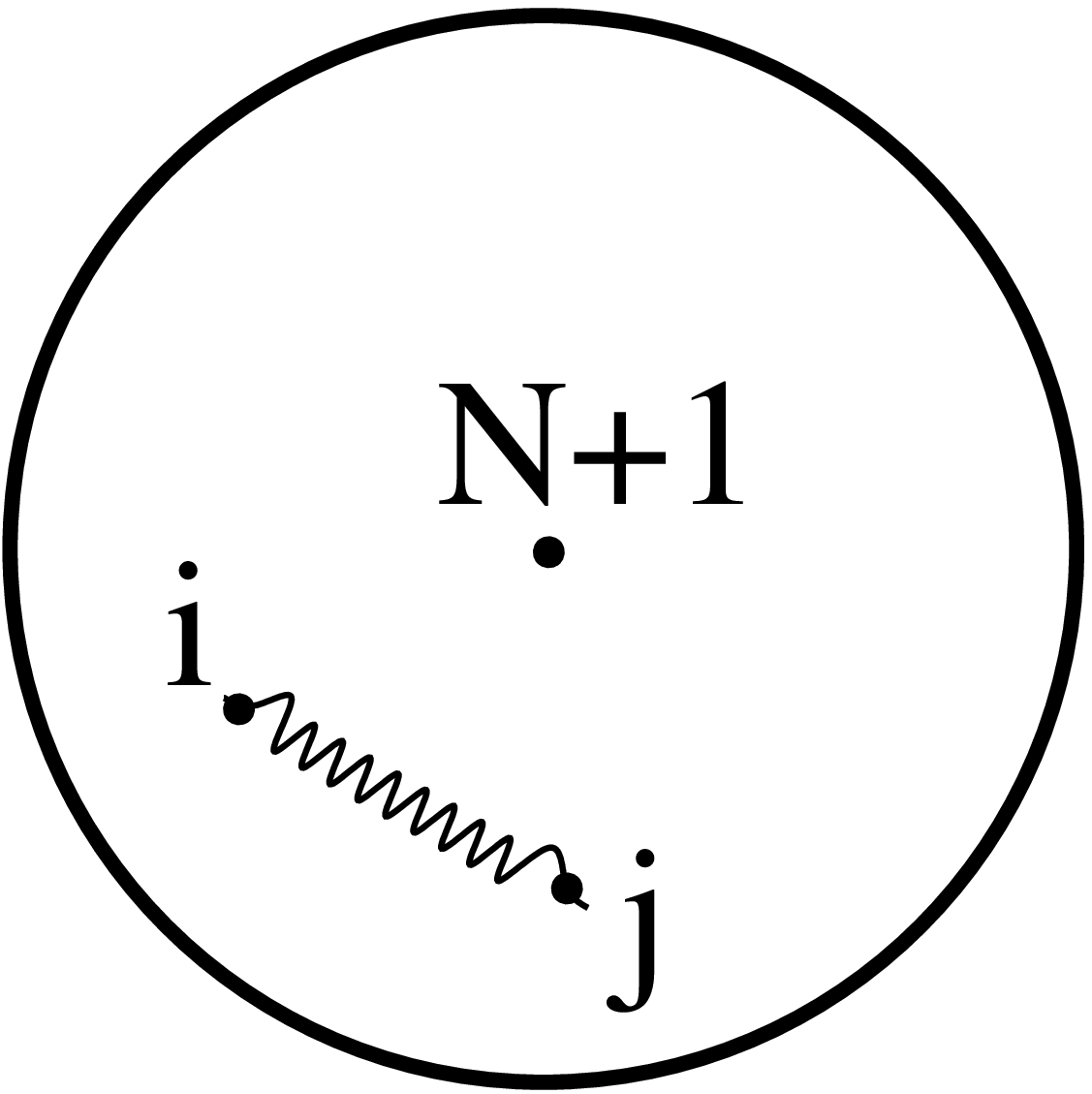}} :=
  \overline{\ln |x_{j}{-}x_{i}|}\ ,
\end{equation}
the $N$-loop integral reads up to first order in $2{-}D$
\begin{eqnarray}
  \label{6.16}
  \int (\det D)^{-d/2} =\mu^{-N \E}
    \left(\frac{1+N}{2^{N}}\right)^{-d/2}\!\!\!\!\!\!\times
\Bigg[1&-&\frac{d}{2}(2{-}D)\left(\frac{2N}{1{+}N}\
      \raisebox{-4.5mm}{\includegraphics[scale=0.1]{./eps/x1_xj.eps}}+\frac{N(N{-}1)}{1{+}N}\ \raisebox{-4.5mm}{\includegraphics[scale=0.1]{./eps/xi_xj.eps}}
  \right)\nonumber\\
  & +& O((2{-}D)^{2}) \Bigg]\ .
\end{eqnarray}
We now turn to an explicit calculation of the diagrams. To first order in
$2{-}D$ we have two contributions. The first (\ref{6.14}) is 
\begin{eqnarray}
  \label{6.18}
  \raisebox{-4.5mm}{\includegraphics[scale=0.1]{./eps/x1_xj.eps}}&=&\int_{x_{i}}
\ln|x_{N{+}1}{-}x_{i}|=D\int_{0}^{1}\frac{\mbox{d}x\
  x^{D}}{x}\ln x 
  =-\frac{1}{D}\longrightarrow -\frac{1}{2}\ ,\quad \textrm{as}\ D\to 2\ .\qquad 
\end{eqnarray}
The second diagram (\ref{6.15}) is 
\begin{displaymath}
  \raisebox{-4.5mm}{\includegraphics[scale=0.1]{./eps/xi_xj.eps}}=\frac{D^{2}}{S^{2}_{D}}\int_{0}^{1}\mbox{d}^{D} x_{i}\ \mbox{d}^{D} x_{j}\
  \ln|\vec x_{i}-\vec x_{j}|\ ,
\end{displaymath}
where we have put vectors to stress that $x_{i}-x_{j}$ is not a
scalar. 
It suffices to evaluate this in $D=2$. We find
\begin{equation}
  \label{6.21}
  \raisebox{-4.5mm}{\includegraphics[scale=0.1]{./eps/xi_xj.eps}}=-\frac{1}{4}\ .
\end{equation}
Combining (\ref{6.16}), (\ref{6.18}) and (\ref{6.21}) we finally obtain for the
renormalized coupling $g$ up to first order in\footnote{The
  expression in parenthesis is the inverse renormalization
$Z_{g}$-factor which relates 
  the bare coupling $g_{0}$ to the renormalized coupling $g$ according to $g Z_{g}\mu^{\varepsilon}=g_{0}$.} $2-D$:
\begin{eqnarray}
  \label{6.23}
  g&=& g_{0}\mu^{-\varepsilon}\ \left(1+\sum_{N=1}^{\infty}\frac{(-g_{0}\
  2^{d/2}\mu^{-\varepsilon})^{N}}{(N{+}1)!(1{+}N)^{d/2}}\right.\times \nonumber\\
  &&\left.\qquad\qquad  \times
  \left[1-(2{-}D)\frac{d}{2}\left(\frac{2N}{1{+}N}\,\raisebox{-4.5mm}{\includegraphics[scale=0.1]{./eps/x1_xj.eps}}+\frac{N(N{-}1)}{1{+}N}\,\raisebox{-4.5mm}{\includegraphics[scale=0.1]{./eps/xi_xj.eps}}\right)+\cdots\right]\right) \nn\\
&=&  g_{0}\mu^{-\varepsilon}\ \left(1+\sum_{N=1}^{\infty}\frac{(-g_{0}\
  2^{d/2}\mu^{-\varepsilon})^{N}}{(N{+}1)!(1{+}N)^{d/2}}
  \left[1+(2{-}D)\frac{d}{2}\left(\frac{N}{1{+}N}+\frac{N(N{-}1)}{4 ( 1{+}N)}\right)+\cdots\right]\right)\ .\qquad 
\end{eqnarray}
This can also be written as
\begin{eqnarray}
g 2^{d/2} &=& - \sum_{N=1}^{\infty } \frac{( -g_{0}2^{d/2}\mu ^{-\E})^{N}}{N!\,N^{d/2}} \nn\\
&& - (2{-}D)\frac{d}{2} \sum _{N=1}^{\infty }  \frac{( -g_{0}2^{d/2}\mu ^{-\E})^{N} (N{-}1)}{N!\,N^{d/2+1}}\nn\\
&&  - (2{-}D)\frac{d}{8} \sum _{N=1}^{\infty }  \frac{( -g_{0}2^{d/2}\mu ^{-\E})^{N} (N{-}1) (N{-}2)}{N!\,N^{d/2+1}}
\ .
\end{eqnarray}
Absorbing a factor of $2^{d/2}$ both in $g$ and $g_{0}$ and defining
the dimensionless coupling $z:=g_{0}\mu ^{-\E}$, we obtain the final result
\begin{eqnarray}
\label{6.26}
g &=& - \sum_{N=1}^{\infty } \frac{( -z)^{N}}{N!\,N^{d/2}} -
(2{-}D)\frac{d}{2} \sum _{N=1}^{\infty } \frac{( -z)^{N}
(N-1)}{N!\,N^{d/2+1}} - (2{-}D)\frac{d}{8} \sum _{N=1}^{\infty }
\frac{( -z)^{N} (N{-}1) (N{-}2)}{N!\,N^{d/2+1}}\nonumber\\
&&+\,O (2-D)^{2}
\ .
\end{eqnarray}
This formula is the starting point for further analysis in the
subsequent sections.

\subsection{Asymptotic behavior of the series}
In the following we    are interested in the limit of large $z$
(strong repulsion), which also is the  scaling behavior of infinitely
large membranes. We therefore need an analytical expression for sums like
(\ref{6.26}) in the limit of large $z$. An example is the leading 
order expression
\begin{equation}
  \label{6.26.1}
  g = -\sum_{N=1}^{\infty}\frac{(-z)^{N}}{N!\ N^{d/2}}\ .
\end{equation}
For the following analysis we need an integral representation of the above series.
We claim that for all $k,d>0$
\begin{equation}\label{6.28}
  \sum_{N=0}^{\infty}\frac{(-z)^{N}}{N!(k+N)^{d/2}}=\frac{1}{\Gamma(\frac{d}{2})}\int_{0}^{\infty}\mbox{d}r\
  r^{d/2-1}\exp[-z\
  \rme^{-r }-k r]\ .
\end{equation}
This can be proven as follows:
\begin{eqnarray*}
  \frac{1}{\Gamma(\frac{d}{2})}\int_{0}^{\infty}\mbox{d}r\
  r^{d/2-1}\exp[-z\,
  \rme^{-r}-k r]&=&\frac{1}{\Gamma(\frac{d}{2})}\sum_{N=0}^{\infty}\frac{(-z)^{N}}{N!}\int_{0}^{\infty}\mbox{d}r\
  r^{d/2-1}\rme^{-(N+k)r}\\
  &=&\frac{1}{\Gamma(\frac{d}{2})}\sum_{N=0}^{\infty}\frac{(-z)^{N}}{N!}\frac{\Gamma(\frac{d}{2})}{(N+k)^{d/2}}\ .
\end{eqnarray*}
This integral-representation is not the most practical for our
purpose. It is better to set $r\to s:=\rme^{-r}$ which yields 
\begin{equation}
  \sum_{N=0}^{\infty}\frac{(-z)^{N}}{N!(k+N)^{d/2}}=\frac{1}{\Gamma(\frac{d}{2})}\int_{0}^{1}\mbox{d}s\, s^{k-1} (-\ln s)^{d/2-1} \rme^{-s z}\ .
\end{equation}
This formula is already very useful for some purposes. It is still
advantageous to make a second variable-transformation $s\to y := sz$, yielding
\begin{equation}
  \sum_{N=0}^{\infty}\frac{(-z)^{N}}{N!(k+N)^{d/2}}=\frac{(\ln z)^{d/2-1}}{\Gamma(\frac{d}{2})z^{k}}\int_{0}^{z}\mbox{d}y\, y^{k-1} \left(1-\frac{\ln y}{\ln z} \right)^{d/2-1} \rme^{-y}\ .
\end{equation}
Finally we remark that we usually have the following combination
\begin{equation}
f_{k}^{d} (z):= z^{k}
\sum_{N=0}^{\infty}\frac{(-z)^{N}}{N!(k+N)^{d/2}}=\frac{(\ln
z)^{d/2-1}}{\Gamma(\frac{d}{2})}\int_{0}^{z}\rmd y\, y^{k-1} \left(1-\frac{\ln y}{\ln z} \right)^{d/2-1} \rme^{-y}\ .
\end{equation}
It satisfies the following simple recursion relation, which is
helpful to calculate the $\beta$-function:
\begin{equation}\label{recursion}
z \frac{\rmd }{\rmd z} f_{k}^{d} (z) = f_{k}^{d-2} (z) \ . 
\end{equation}
From (\ref{6.28}) $f_k (z)>0$ for all $k,d>0$ and the behavior for large $z$ is obtained by expanding
$\left(1-\frac{\ln y}{\ln z} \right)^{d/2-1}$ for small $\frac{1}{\ln z}$
\begin{eqnarray}
f_{k}^{d} (z) &=\ds \frac{(\ln z)^{d/2-1}}{\Gamma(\frac{d}{2})} \Bigg[&
\int_{0}^{\infty } \rmd y\, y^{k-1} \rme^{-y} \nonumber \\
&& -\frac{1}{\ln z}\left(\frac{d}{2}-1 \right)\int_{0}^{\infty }
\rmd y\, y^{k-1}\ln y\, \rme^{-y}\nonumber \\
&& +\, O  \left(\frac{1}{(\ln z)^{2}} \right)
 \Bigg]\nonumber \\
&\ds +\, O (\rme^{-z})\ \ \hfill&
\ .
\end{eqnarray}
The result is 
\begin{equation}
f_{k}^{d} (z) =\frac{(\ln z)^{d/2-1}\Gamma (k)}{\Gamma(\frac{d}{2})}
\left(1- \frac{1}{\ln z} \frac{d{-}2}{2} \frac{\Gamma' (k)}{\Gamma
(k)} + \dots  \right)
\ .
\end{equation}
For later reference, we note  a list of useful formulas ($\gamma
=0.577216\dots $ is the Euler-constant)
\begin{eqnarray}
f_{1}^{d} (z) &=& \frac{(\ln z)^{d/2-1}}{\Gamma
(\frac{d}{2})}+\frac{\gamma (\ln  z) ^{d/2-2}}{\Gamma (\frac{d-2}{2})}
+\frac{( 6 \gamma ^{2}+\pi ^{2}) (\ln  z)^{d/2-3}}{12 \Gamma (\frac{d-4}{2})}+ \dots \\
f_{2}^{d} (z) &=& \frac{(\ln z)^{d/2-1}}{\Gamma
(\frac{d}{2})}+\frac{(\gamma-1) (\ln z)^{d/2-2}}{\Gamma (\frac{d-2}{2})}
+\frac{( 6 \gamma (\gamma -2) +\pi ^{2}) (\ln z)^{d/2-3}}{12 \Gamma (\frac{d-4}{2})}+ \dots\\
f_{3}^{d} (z) &=& \frac{2 (\ln  z)^{d/2-1}}{\Gamma
(\frac{d}{2})}+\frac{(2\gamma-3) (\ln z)^{d/2-2}}{\Gamma (\frac{d-2}{2})}
+\frac{( 6 \gamma^{2} -18 \gamma +6 +\pi ^{2}) (\ln z)^{d/2-3}}{6 \Gamma (\frac{d-4}{2})}+ \dots\ .\nn\\
\end{eqnarray}
Note that with the above notations, the relation (\ref{6.26})
expressing $g$ as a function of $z$ becomes
\begin{equation}\label{star}
g (z) = f_{1}^{{d+2}} (z) - (2-D)\frac{d}{2} f_{2}^{d+4} (z) +
(2-D)\frac{d}{8}f_{3}^{d+4} (z) + O (2-D)^{2}\ .
\end{equation}

In the next two sections we shall present two formalisms which give 
results for $\omega $ at next to leading order. It will turn out that
the results catch qualitatively correctly the cross-over to $D=1$;
however they differ in the numerical values. This is a reflection of
the fact that   the  $( 2{-}D)$-expansion is not
systematic. We expect that working on a manifold of finite size and using
the exact correlation function on this manifold would lead to a
systematic expansion. Work is 
in progress \cite{PinnowWieseProgress} to check this hypothesis.

\subsection{The RG-functions in the bare coupling}
\begin{figure}[t]
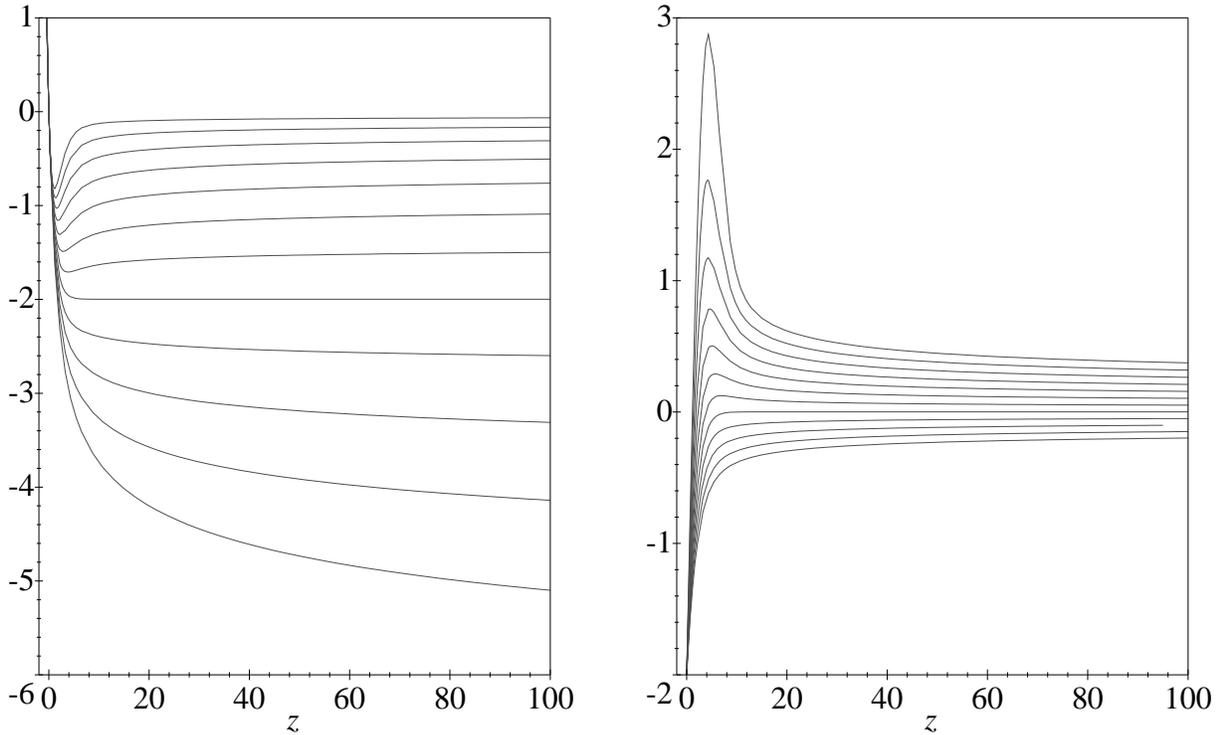

\noindent
\epsfxsize=0.47\textwidth\epsfbox{./eps/betaN.epsi}
\hfill 
\epsfxsize=0.47\textwidth\epsfbox{./eps/omegaN.epsi}
\vspace{-1cm}
    \caption{$\beta$-function (left) and $\omega$ (right) as functions of the dimensionless
      bare coupling $z$ for different dimensions $d$: $d=0.25,\ 0.5,\
      0.75,\dots,\ 3$ (from top to bottom, respectively). Note, that $\beta$
      always has a fixed point at $z{=}0$ with
      $\omega(0){=}-\varepsilon$. Furthermore, there is a fixed point for
      $0{<}d{<}2$ at $z{=}\infty$ with $\omega(\infty){=}0$, the latter
      remaining true for $d{\geq}2$. In $d{=}2$ $\beta$ tends to $-\varepsilon$
      as $z\to \infty$. Above, $\beta$ diverges.}
    \label{f6.2}
\end{figure}

The renormalization group $\beta$-function is defined as
\begin{equation}
\beta (g):= \mu \frac{\rmd }{\rmd \mu }\lts_0 g\ .
\end{equation}
Since it is hard to invert the function $g (z)$, it is advantageous to
study  $\beta $ as function of the bare coupling $z$
\begin{equation}
  \label{fb.7}
\beta (g)\equiv \beta (z) = -\E z \frac{\rmd }{\rmd z}g\ .
\end{equation}
To first order in $2-D$, we find from (\ref{star}) using
(\ref{recursion}) 
\begin{equation}
  \label{fb.8}
  \beta (z) =  -\varepsilon\left
  [ f_{1}^{d} (z) + (2-D)\left( -\frac{d}{2} f_{2}^{d+2}(z)+\frac{d}{8}
  f_{3}^{d+2} (z)\right)+ O (2-D)^{2}\right]\ .
\end{equation}
Inserting the asymptotic expansions from the last section, this 
becomes
\begin{equation}\label{fb.8.a}
  \beta (z) =  -\frac{\varepsilon}{\Gamma (\frac{d}{2})}\left[
(\ln z)^{d/2-1} - \frac{2-D}{2} ( \ln z)^{d/2} +\dots 
\right]
\ ,
\end{equation}
where  omitted terms have either one more power of $(2-D)$ or
$1/\ln z$. 
As shown in section \ref{Model and physical observables}, universal
properties are given by the correction to scaling exponent
$\omega$, reading 
\begin{equation}
  \label{fb.9}
  \omega(z):=\frac{\rmd \beta (g)}{\rmd g}=\frac{\rmd \beta (z)}{\rmd z}\frac{1}{\frac{\rmd g}{\rmd z}}=-\frac{\varepsilon z}{\beta (z)}\frac{\mbox{d}\beta(z)}{\mbox{d}z}\ .
\end{equation}
Let us first study  $D=2$. $\beta (z)$ and $\omega (z)$ are plotted on
figure \ref{f6.2}. We see that for $d<2$ the $\beta$-function becomes
0 for $z\to\infty$. For $d>2$, $\beta (z)$ has no zero, but the flow
of $g$ is still to infinity. 
In both cases $\omega $ is given by $\omega =\lim_{z\to\infty}\omega (z)$.
Taking only the leading term in the $(2-D)$-expansion of $\beta (z) $,
$\omega (z)$ is given by  
\begin{equation}
\omega (z) = -\varepsilon \frac{z\frac{\rmd}{\rmd z}f_{1}^{d}(z)}{f_{1}^{d}(z)}= \frac{d-2}{2 \ln z} \to  0\ .
\end{equation}
Let us now  take into account the first order in $2-D$. Then the fixed
point of (\ref{fb.8.a}) is at the finite value
\begin{equation}\label{fb.8.b}
\ln z^{*} =  \frac{2}{2-D}\ . 
\end{equation}
Next, when trying to calculate $\omega (z)$, we face the following
problem: Since we truncated the series for $g$ at order one in $2-D$,
$\beta' (z)$ does not vanish at the fixed point, i.e.\ the zero of
$\beta (z)$. This might lead to the conclusion that $\omega (z^{*})$
is always $\infty $, an absurd result. However since this is a consequence
of the truncation of the series, we follow the strategy also to expand
the denominator of $\omega (z)$ in powers of $2-D$. From (\ref{fb.8})
and (\ref{fb.9}) we obtain:
\begin{eqnarray}
  \label{fb.12}
  \omega(z)=-\varepsilon\left[\frac{z\frac{\rmd}{\rmd
z}f_{1}^{d}(z)}{f_{1}^{d}(z)}\right. &+& (2-D)\frac{z\frac{\rmd }{\rmd
z}\left[
-\frac{d}{2}f_{2}^{d+2}(z)+\frac{d}{8}f_{3}^{d+2}(z)\right]}{f_{1}^{d}(z)}\nonumber\\
    &+&(2-D)\left.\frac{\left[\frac{d}{2}f_{2}^{d+2}(z)-\frac{d}{8}f_{3}^{d+2}(z)\right]z\frac{\rmd }{\rmd z} f_{1}^{d}(z) }{f_{1}^{d}(z)^{2}}\right]
\ .
\end{eqnarray}
Inserting the asymptotic series, we find
\begin{equation}
\omega (z) = -\E \frac{d-2}{2\ln z}+ \E\frac{2-D}{2}\ .
\end{equation}
Inserting the fixed point (\ref{fb.8.b}), we arrive at
\begin{equation}
  \label{fb.13}
\omega (z^{*}) = \E \,(2-D)\, \frac{4-d}{4}\ .
\end{equation}
This should be checked against the exact result in $D=1$, which 
reads
\begin{equation}
\omega (D{=}1) = \E \ .
\end{equation}
Thus for $d\to0$ our resummation-procedure gives the exact and 
for $d>0$ an approximative result. Before analyzing the  validity of 
the procedure, let us turn to the second scheme, namely the
calculation in the renormalized coupling.

\subsection{Calculation in the renormalized coupling}
We start from $g (z)$ given in (\ref{star}). For large $z$ this can be
approximated by
\begin{equation}
  \label{fr.2}
  g(z)=\frac{(\ln
  z)^{d/2}}{\Gamma(\frac{d}{2}{+}1)}-(2-D)\frac{d}{4}\frac{(\ln
  z)^{d/2+1}}{\Gamma(\frac{d}{2}{+}2)}\ ,
\end{equation}
where we only retained the leading term from each series. Also note that this
formula is only valid for $d<2$. For $d\ge2$ additional additive
constants have to be added. Using (\ref{fr.2}), we can 
write $\ln z$ as a function of $g$. To first order in $2-D$, this reads
\begin{equation}
\ln z  = \tilde g^{2/d} + \frac{2-D}{2+d} \, \tilde g^{4/d}\ ,
\qquad 
\tx \tilde g:=g\,\Gamma(\frac{d}{2}+1)
\ .
\end{equation} 
\begin{figure}[t]
  \begin{center}
    \vspace{1cm}
    \setlength{\unitlength}{1cm}
    \begin{picture}(7,5)
    \put(0,0){\raisebox{0mm}{\includegraphics[scale=0.8]{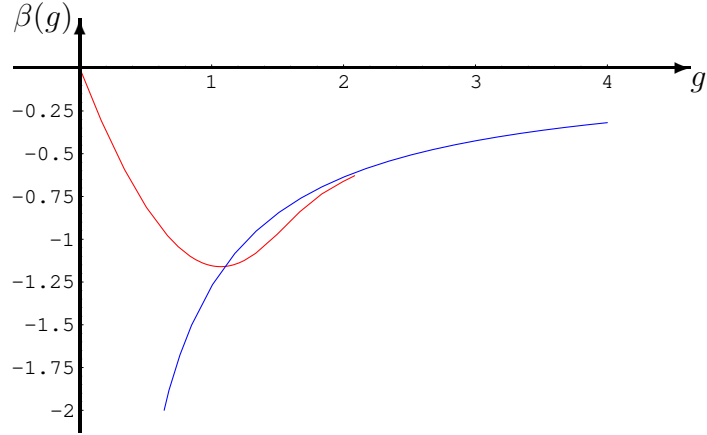}}}
    \thicklines
    \put(0,4.65){\vector(1,0){9.0}}
    \put(0.88,-0.2){\vector(0,1){5.5}}
    \put(9.0,4.4){$g$}
    \put(-0.0,5.2){$\beta(g)$}
      \end{picture}
    \end{center} 
\caption{$\beta-$function in terms of the
    renormalized coupling $g$  truncated at order 160, Pade-resummed,
and plotting 
only that part for which the truncated series converges. (This can
e.g.\ be tested by taking away the last few terms of the series.) This
is  compared to the  asymptotic behavior (\ref{fr.10})
    (proportional to $1/ g$ for large $g$).  $d$ is set to $1$,
and we used the diagonal (80,80)-Pade approximant, which was find to
converge best. (The non-resummed expression starts to diverge already at
$g\approx 1.8$ at this order.)} 
    \label{betapa}
\end{figure}
We now write the $\beta$-function in terms of $\tilde g$. Starting from
(\ref{fb.8.a})
we obtain
\begin{equation}
  \label{fr.10}
  \beta(\tilde g)=-\frac{\varepsilon}{\Gamma(\frac{d}{2})}\left[{\tilde
  g^{1-2/d}}-\frac{2 (2{-}D)}{2{+}d}\tilde
  g\right]\ .
\end{equation}
Fixed points are at the zeros of the $\beta$-function. The non-trivial
one is at
\begin{eqnarray}
  \label{fr.12}
  \tilde g^{*}=\left[\frac{2{+}d}{2 (2{-}D)}\right]^{d/2}\ .
\end{eqnarray}
The correction to scaling exponent $\omega$ is simply obtained by evaluating
the derivative of $\beta$ with respect to $g$ at the fixed point. In terms
of $\tilde g$  this is
\begin{eqnarray}
  \label{fr.13}
  \omega(\tilde g)&=&\frac{\rmd \beta (g)}{\rmd g}=\frac{\rmd \beta(\tilde
  g)}{\mbox{d}\tilde g}\frac{\rmd \tilde g}{\rmd g}=\Gamma({\tx\frac{d}{2}+1})\, \frac{\mbox{d}\beta(\tilde
  g)}{\mbox{d}\tilde g}\nonumber\\
  &=&-\varepsilon\left[\left(\frac{d}{2}-1\right)\tilde g^{-2/d}-\frac{d (2{-}D)}{2{+}d}\right]\ .
\end{eqnarray}
 Inserting the fixed point
$\tilde g^{*}$ from (\ref{fr.12}) we find
\begin{equation}
  \label{fr.14}
  \omega(\tilde
  g^{*})=\varepsilon\,\frac{2 (2{-}D)}{2{+}d}\ .
\end{equation}
Again, this is quite close to the exact result 
 $\omega (D{=}1) =\E$ in $D=1$, and in effect exact in $D=1$ and $d=0$.

Let us finally point out that in the limit $D=2$ the true asymptotic
behavior of the $\beta-$function in terms of the renormalized coupling
$g$ is obtained from the completely summed series (\ref{6.26.1})
leading to (\ref{fr.10}) for large $g$. Conversely, if one tries to
invert (\ref{6.26.1}) and truncates it taking only a finite number of
orders into account, it is at least possible to reach the asymptotic
regime -- however, for large enough $g$ the truncated $\beta-$function
wildly oscillates and thus strongly deviates from the true
behavior. In figure (\ref{betapa}) the Pade-resummed truncated
$\beta-$function up to order $g^{160}$ in $d=1$ is compared with the exact,
asymptotic flow-function. One notices that the truncated
$\beta$-function eventhough improved throuh a Pade-Resummation hardly
gets into touch with the asymptotic regime. The same applies to the
slope-function $\omega(g)$, which is not shown in fig. (\ref{betapa}).

Note that the above arguments suggest quite intuitively the behavior
of the exact $\beta $-function in $1<D<2$. Whereas for $D=1$, the 
$\beta $-function is a parabola, and for $D=2$ it decays like a
power-law for large $g$ at least as long as $d<2$, the $\beta$-function
for values of $D$ between these two extremes should cut the axes
$\beta (g)=0$ at a finite value of $g$, which for $D\to 2$ wanders off 
to infinity, thus by continuity forcing the exponent $\omega $ to go 
to 0 for $D\to 2$. 
\begin{figure}[t]
\centerline{\fig{0.5\textwidth}{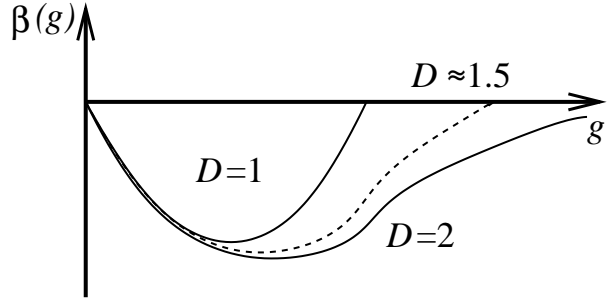}}
\caption{Qualitative behavior for the $\beta $-function in $D=1$, $D=2$
and result anticipated for $D\approx 1.5$.}
\label{guessbeta}
\end{figure}

\subsection{An instructive case: $d=0$}
Setting $d=0$ in the leading order series (\ref{6.26.1}) allows for a
simple analytic expression for $g$ as a function of $z$ and its inversion
\begin{eqnarray}\label{fd0.0}
  g (z)&=&1-\rme^{-z} \\
z (g)&=& -\ln (1-g) \ .
\end{eqnarray}
This gives
\begin{eqnarray}\label{fd0.1}
  \beta(g)&=& \varepsilon(1-g)\ln(1-g) \\
  \omega(g)&=&-\varepsilon(1+\ln(1-g))\ .
\end{eqnarray}
Fixed points are $g^{*}=0 $ and $g^{*}=1$ with $\omega(g^{*}{=}0)=-\varepsilon$
and $\omega(g^{*}{=}1)=\infty$.  Note that the latter result is
an artifact of $d=0$. 

\section{Discussion and conclusion}\label{conclusion}
In this article, we have discussed the perturbation expansion of 
a $D$-dimensional elastic manifold interacting via a $\delta
$-interaction with a  fixed point in embedding space. This calculation
can be done exactly for $D=1$, but becomes non-trivial for $D\neq 1$. 
Interestingly and quite surprisingly, it simplifies considerably in
the limit of $D\to2$, if one decides to work at finite $\E $. In that limit
we were able to obtain an explicit expression for the renormalized
coupling as function of the bare one, in terms of a non-trivial
series. Analysis of this series shows that the fixed point lies at
infinity in the bare coupling, a limit in which we were able to derive an
asymptotic expansion for the renormalized coupling as a function of
the bare one. This yields a vanishing exponent $\omega $ in the
limit of 
$D\to 2$. Also it is important to note that this result is completely
independent of the regularization procedure.
  This does no longer hold
true  beyond the leading order, which should be accessible to 
an expansion in $(2-D)$. We here constructed its first order in a
specific regularization scheme. While this reproduces qualitatively
correctly the known results in $D=1$ (and even exactly for $d=0$), it
also  shows by its dependence 
on the renormalization scheme (here working with the $\beta
$-function either in terms of the renormalized or bare coupling) that
this expansion in $(2-D)$ is  not systematic. Even though
experience with this new kind of expansion is still lacking, this is
likely to be caused by the use of  a hard cutoff in position space while
working with the correlator of an infinite membrane. It seems that
only in an $\E$-expansion this procedure is systematic. Work is in
progress \cite{PinnowWieseProgress} to study the model on a sphere or
torus of finite size, such 
that no further infrared cutoff is necessary. 
This should lead to exact results beyond the leading order and provide
for the crossover between polymers and membranes. 

While results for the pinning problem are interesting in its own, the
main motivation is certainly to obtain a better understanding of 
self-avoiding polymerized membranes. This model is well-behaved
physically, since  its fractal dimension $d_{f}$ is bounded by the
dimension of the embedding space $d_{f}<d$.  Preliminary studies
\cite{PinnowWieseProgress} indicate that  this problem can also
be attacked by the methods developed in this article. This would be
very welcome to check the 2-loop calculations
\cite{WieseDavid1997,DavidWiese1996} and the large-order behavior
\cite{DavidWiese1998} on one side and numerical results (e.g.\
\cite{BowickTravesset2001})  on the other.  


\begin{appendix}
\section{Universal $1/r\ $- repulsion law for polymers}
We have shown  in section \ref{wall-force} by a scaling argument that as long as  $r\ll L^{\nu}$, the restricted partition function of
a manifold pinned at one of its internal points scales as
$
  \mathcal{Z}_{\infty}(r/L^{\nu})\sim
  (r/L^{\nu})^{\theta}$, with the contact-exponent given
by (\ref{2.19}). It is instructive to prove this  for the special case
of polymers in $d=1$. According to (\ref{2.0}) and (\ref{2.1}) this
corresponds to a polymer in $3d$-space interacting with a $\delta$-potential
on a  plane.

Let us start from the (bulk) polymer propagator being defined as the
conditional probability of finding the internal point
$s^{\prime}$ at the position $r_{f}$ when starting in $s$ at $r_{i}$.
The propagator  can be written as a functional integral for a restricted
partition function of the free chain according to
\begin{eqnarray}
  \label{a.3}
  \mathcal{Z}_{s,s^{\prime}}(r_{i}|r_{f}) & =&  \int \mathcal{D}[r]\
  \tilde \delta(r(s){-}r_{i})\tilde \delta(r(s^{\prime}){-}r_{f})\
  e^{-\mathcal{H}[r]}\nonumber\\
  & =& \ \int_{k}
  e^{-k^{2}|s^{\prime}{-}s|{+}ik(r_{f}{-}r_{i})}\ =\ |s^{\prime}{-}s|^{-d/2}\
  e^{-\frac{(r_{f}{-}r_{i})^{2}}{4|s^{\prime}{-}s|}}\ .
\end{eqnarray}
The propagator (\ref{a.3}) posesses the Markov property
\begin{equation}
  \label{a.4}
  \mathcal{Z}_{s,s^{\prime \prime}}(r_{i}|r_{f}) = \int \mbox{d}^{d}r_{m}\
  \mathcal{Z}_{s,s^{\prime}}(r_{i}|r_{m})\ \mathcal{Z}_{s^{\prime},s^{\prime
  \prime}}(r_{m}|r_{f})\ .
\end{equation}
In the following we are interested in the restricted partition function
(\ref{a.3}) in presence of a short-range interaction modeled through a
$\delta$-potential, which is situated at the origin of the embedding
space. The restricted partition function in the presence of the
interaction will be denoted by
$\mathcal{Z}^{g_{0}}_{s,s^{\prime}}(r_{i}|r_{f})$. We consider the case,
where the chain is fixed at its ends, such that $s{=}0$ and
$s^{\prime}{=}L$. We furthermore switch to a grand-canonical ensemble and
denote
\begin{equation}
  \label{a.5}
  \mathcal{Z}^{g_{0}}_{\tau}(r_{i}|r_{f}) :=
  \int_{0}^{\infty }\rmd L \, \mathcal{Z}^{g_{0}}_{0,L}(r_{i}|r_{f})\ e^{-\tau L}\ ,
\end{equation}
$\tau$ being some chemical potential. $\mathcal{Z}^{g_{0}}_{\tau}(r_{i}|r_{f})$ can be expanded in a
perturbation series in the coupling $g_{0}$ according to
\begin{equation}
  \label{a.6}
  \mathcal{Z}^{g_{0}}_{\tau}(r_{i}|r_{f})\ =\
  \sum_{n=0}^{\infty}(-g_{0})^{n}\mathcal{Z}^{g_{0},n}_{\tau}(r_{i}|r_{f})\ .
\end{equation}
The $\mathcal{Z}^{g_{0},n}_{\tau}(r_{i}|r_{f})$ are 
\begin{eqnarray}
  \label{a.7}
  \mathcal{Z}^{g_{0},n}_{\tau}(r_{i}|r_{f}) &:=& \int\limits_{0}^{\infty}\mbox{d}L\
  \int\limits_{0{<}x_{1}{<}\dots {<}x_{n}{<}L}
  \mathcal{Z}_{0,x_{1}}(r_{i}|0)\ \mathcal{Z}_{x_{1},x_{2}}(0|0)\times \dots
  \nonumber\\
  &&\dots \times \mathcal{Z}_{x_{n{-}1},x_{n}}(0|0)\
  \mathcal{Z}_{x_{n},L}(0|r_{f})\ \rme^{-\tau L}\nonumber \\
&=&\! \left[ \int_{0}^{\infty}\rmd x\ \rme^{-x \tau} 
 \mathcal{Z}_{0,x} (r_{i},0) \right]\!\!
\left[ \int_{0}^{\infty}\rmd x\ \rme^{-x \tau} 
 \mathcal{Z}_{0,x} (0,0) \right]^{n{-}1}\!
\left[ \int_{0}^{\infty}\rmd x\ \rme^{-x \tau} 
 \mathcal{Z}_{0,x} (0,r_{f}) \right]\!\! \nn\\
&=&  \mathcal{Z}_\tau ^{0} (r_{i}|0)\,  \mathcal{Z}_\tau ^{0} (0|0)^{n-1}\,  \mathcal{Z}_\tau ^{0} (0|r_{f}) \ ,
\end{eqnarray}
where we have used that after integration over $L$,
the integrals factorize; we further have defined
\begin{eqnarray}
\mathcal{Z}_{\tau } (r_{i}|r_{f}) &=& \int_{0}^{\infty } \rmd s
\mathcal{Z}_{0,s} (r_{i}|r_{f})   \rme^{-\tau s} \nn\\
\label{a:9}
&=&  \int_{0}^{\infty } \rmd s \int_{k}\rme^{- ( k^{2}+\tau )s+ ik
(r_{f}-r_{i})} =\int_{k}\frac{\rme^{ik (r_{f}-r_{i})}}{k^{2}+\tau }\ .
\end{eqnarray}
In $d{=}1$ the integration over the momenta $k$ in
(\ref{a:9}) leads to 
\begin{equation}
  \label{a.10}
  \mathcal{Z}_{\tau}(r_{i}|r_{f})_{|d{=}1}\ =\ \sqrt{\frac{\pi}{\tau}}\
  e^{-\sqrt{\tau}|r_{f}{-}r_{i}|}\ .
\end{equation} 
Let us now lock at what terms appear in the $n$-th order coefficient in
(\ref{a.7}): From the initial point of the chain a factor
$\mathcal{Z}_{\tau}(r_{i}|0)$ is contributed, while the final point is coming
up with $\mathcal{Z}_{\tau}(0|r_{f})$. Furthermore, the internal points give
$n{-}1$ powers of $\mathcal{Z}_{\tau}(0|0)$, where
\begin{equation}
  \label{a.11}
  \mathcal{Z}_{\tau}(0|0) = \sqrt{\frac{\pi}{\tau}}\ .
\end{equation}
The $(n{=}0)$-coefficient is special: Clearly, one has
\begin{equation}
  \label{a.12}
  \mathcal{Z}^{g_{0},0}_{\tau}(r_{i}|r_{f}) =
  \mathcal{Z}_{\tau}(r_{i}|r_{f})\ .
\end{equation}
Inserting (\ref{a.11})-(\ref{a.12}) into (\ref{a.7}) we obtain for (\ref{a.6}):
\begin{eqnarray}
  \label{a.13}
  \mathcal{Z}^{g_{0}}_{\tau}(r_{i}|r_{f}) &=& \sqrt{\frac{\pi}{\tau}}\
  \left[\rme^{-\sqrt{\tau}|r_{f}{-}r_{i}|}+ \rme^{-\sqrt{\tau}(|r_{f}|{+}|r_{i}|)}\
  \sum_{n{=}1}^{\infty}  \left(-g_{0}\sqrt{\frac{\pi}{\tau}}\right)^{n}\right]\nn\\
&=& \sqrt{\frac{\pi}{\tau}}\
  \left[\rme^{-\sqrt{\tau}|r_{f}{-}r_{i}|}- \rme^{-\sqrt{\tau}(|r_{f}|{+}|r_{i}|)}\
 \frac{1}{1+\frac{1}{g_0}\sqrt{\frac{\tau}{\pi}}}   \right]
\ .
\end{eqnarray} 
The limit of  $g_{0}\to \infty $ can be taken and leaves us with the
beautiful result
\begin{equation}
  \label{a.15}
  \mathcal{Z}^{\infty}_{\tau}(r_{i}|r_{f}) = \sqrt{\frac{\pi}{\tau}}\
  \left(\rme^{-\sqrt{\tau}|r_{f}{-}r_{i}|}-\rme^{-\sqrt{\tau}(|r_{f}|{+}|r_{i}|)}\right)\ .
\end{equation}
This is the propagator of a scalar field with Dirichlet boundary
conditions. It has applications when studying critical phenomena of
e.g.\ an Ising magnet in 
 half-space \cite{DiehlInDombGreen,EisenrieglerKremerBinder1982}. Performing the inverse Laplace-transformation of (\ref{a.15}) we arrive at
\begin{equation}
  \label{a.16}
  \mathcal{Z}^{\infty}_{L}(r_{i}|r_{f}) = 
  \frac{1}{L^{1/2}}\left(\rme^{-\frac{(r_{f}{-}r_{i})^{2}}{4|L|}}-\rme^{-\frac{(|r_{f}|{+}|r_{i}|)^{2}}{4|L|}}\right)\ .
\end{equation}
We  conclude that a $\delta$- potential acting in a
hyper-plane suppresses all configurations which penetrate or touch it,
when taking its amplitude to infinity.\\
Finally, in order to prove the universal repulsion law, 
 we start from (\ref{a.16}) and
integrate over all final positions $r_{f}$:
\begin{eqnarray}
  \label{a.17}
  \mathcal{Z}^{\infty}_{L}(r_{i}) = \int\limits_{-\infty}^{\infty}\mbox{d}r_{f}\
  \mathcal{Z}^{\infty}_{L}(r_{i}|r_{f}) = 2\sqrt{\pi}\left(1-\mbox{erf}
  \frac{|r_{i}|}{2\sqrt{L}}\right)\ ,
\end{eqnarray}
$\mbox{erf}$ denoting the error function. For small arguments we have
\begin{equation}
  \label{a.18}
  1-\mbox{erf}(x)\ \sim \ x\ .
\end{equation}
Furthermore, $\mathcal{Z}^{\infty}_{L}(r_{i})$ denotes the restricted
partition function of a chain  pinned at one of its ends in $r_{i}$. To
obtain $\mathcal{Z}_{\infty}(r/L^{\nu})$ in (\ref{2.17}) we have to
evaluate:
\begin{equation}
  \label{a.19}
  \mathcal{Z}_{\infty}(r/L^{\nu}) = \frac{1}{L}\int_{0}^{L}\mbox{d}s\
  \mathcal{Z}^{\infty}_{s}(r)\mathcal{Z}^{\infty}_{L{-}s}(r)\ \sim \ \left(\frac{r}{\sqrt{L}}\right)^{2}
\end{equation}
according to (\ref{a.18}) in the scaling regime $r{\ll}L^{1/2}$. We
thus find that $\theta =2$, which confirms (\ref{2.19}) for $D=1$ and $d=1$.
The above proof can be extended to arbitrary dimensions $0{<}d{<}2$, where
$\varepsilon{>}0$ for polymers. Then,
$\theta = 4-2d$. 

It would be nice to make the same arguments for membranes. 
However, the  proof is based on a drastic simplification, which only occurs in
$D{=}1$ and shows up in the factorizability of loop diagrams as in
(\ref{a.7}). This has no  extension to manifolds of internal dimension $D{>}1$.
\section{Conformal mapping of the sectors}
\label{conformal} 
\begin{figure}[t]
\begin{center}
    \leavevmode
    { \setlength{\unitlength}{1cm}
    \begin{picture}(8,5)
    \put(0,0){\raisebox{0mm}{\includegraphics[scale=0.5]{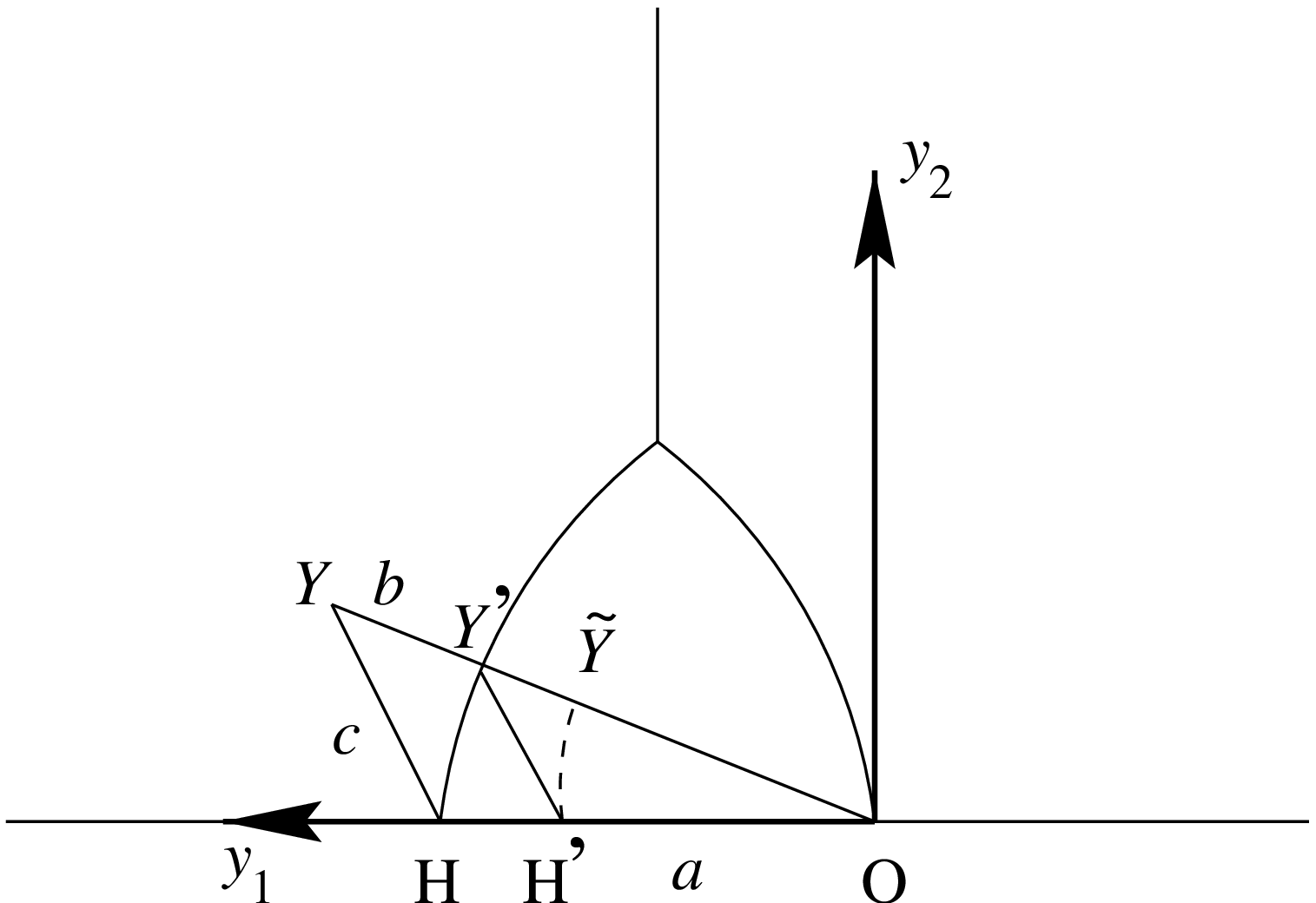}}} 
    \put(3.8,1.5){$\mathcal{A}$}
    \put(1.8,3.5){$\mathcal{B}$}
    \put(5.8,3.5){$\mathcal{C}$}
      \end{picture} }
\end{center}
    \caption{The sectors $\mathcal{A},\mathcal{B}$ and $\mathcal{C}$}
    \label{fb.0}
\end{figure}

In order to calculate the two-loop diagram efficiently, one wants to
write it as an integral over a finite domain only.  To do so we need the technique of conformal
mapping of the sectors, which also serves for analytically continuing the
measure of integration to internal dimensions $D<1$. This  technique has
been extensively used and well documented in the context of  self-avoiding tethered membranes
\cite{WieseDavid1995,WieseDavid1997,WieseHabil}, but we repeat the
presentation here for completeness.\\  
Generally, in evaluating the three-point divergences we need to
integrate over some domain in the upper half-plane (see
\ref{4.18}). The measure of integration reads
\begin{equation}
  \label{b.0}
  \int_{y}\ =\
  \frac{S_{D{-}1}}{S_{D}}\int\limits_{-\infty}^{\infty}\mbox{d}y_{1}\
  \int\limits_{0}^{\infty}\mbox{d}y_{2}\ (y_{2})^{D{-}2}
\end{equation}
and the integrand is a function $f$ of the three distances $a,b$ and $c$
between the points $O=(0,0)$, $H=(-L,0)$ and $Y=(y_{1},y_{2})$, given
explicitly by (see fig.\ \ref{fb.0})
\begin{equation}
  \label{b.1}
  a=L=\textrm{fixed},\quad b=\sqrt{(y_{1})^{2}+(y_{2})^{2}}\ ,\quad
  c=\sqrt{(L{-}y_{1})^{2}+(y_{2})^{2}}\ .
\end{equation}%
In our problem $f$ is homogeneous, of degree $\lambda$, but not necessarily
symmetric:
\begin{equation}
  \label{b.2}
  f(\kappa a,\kappa b,\kappa c)\ =\ \kappa^{-\lambda}f(a,b,c)\ .
\end{equation}
Let us
now explicitly show, how sectors can be mapped onto each other. Specializing
to the mapping $\mathcal{B}\to \mathcal{A}$ we choose a coordinate system as
given in figure (\ref{fb.0}). The mapping $\mathcal{B}\to \mathcal{A}$ is
mediated by a special conformal transformation, the inversion with respect to
the circle around $O$ and radius $L$. In complex coordinates this is
($\bar Y$ is the complex conjugate of $Y$)
\begin{equation}
  \label{b.3}
  Y \longrightarrow \tilde Y = \frac{L^{2}}{\bar Y} = Y\,\frac{L^{2}}{b^{2}}\ \Leftrightarrow \ Y =
  \tilde Y \frac{L^{2}}{\tilde b^{2}}
\end{equation}
such that
\begin{eqnarray}
  \label{b.4}
  \left(
  \begin{array}{cc}
      y_{1} \\
      y_{2}
    \end{array}\right)\ \longrightarrow \  \left(
  \begin{array}{cc}
      \tilde y_{1} \\
      \tilde y_{2}
    \end{array}\right)\ =\ \left(
  \begin{array}{cc}
      \frac{y_{1}L^{2}}{b^{2}} \\
      \frac{y_{2}L^{2}}{b^{2}}
    \end{array}\right)\ \Leftrightarrow \ \left(\begin{array}{cc}
      y_{1} \\
      y_{2}
    \end{array}\right)\ =\ \left(
  \begin{array}{cc}
      \frac{\tilde y_{1}L^{2}}{\tilde b^{2}} \\
      \frac{\tilde y_{2}L^{2}}{\tilde b^{2}}
    \end{array}\right)\ .
\end{eqnarray}
This change of coordinates gives a Jacobian for the measure (\ref{b.0})
\begin{equation}
  \label{b.5}
  \left(\frac{L^2}{\tilde b^{2}}\right)^{D}\ .
\end{equation}
It can  easily be seen that the
mapping $\mathcal{B}\to \mathcal{A}$ is one to one: First, be $Y{=}(y_{1},y_{2})\in
\mathcal{B}$, that is $b>L$ and $b>c$. Then,
\begin{equation}
  \label{b.6}
  \tilde c^{2}\ =\ ((\tilde y_{1}{-}L)^{2}+(\tilde y_{2})^{2})\ =\
  \frac{L^{4}-2y_{1}L^{3}+b^{2}L^{2}}{b^{2}}\ =\ L^{2}\frac{c^{2}}{b^{2}}<L^{2}
\end{equation}
and
\begin{equation}
  \label{b.7}
  \tilde b^{2}\ =\ L^{2}\frac{L^{2}}{b^{2}}<L^{2}\ .
\end{equation}
Since $\tilde b^{2},\tilde c^{2}<L^{2}$, $\tilde Y \in \mathcal{A}$. Second, for the
inverse mapping, $Y(\tilde Y\in
\mathcal{A})\in \mathcal{B}$ is checked as follows:
\begin{equation}
  \label{b.8}
  c^{2}\ =\ b^{2}+L^{2}-2y_{1}L\ =\ b^{2}+L^{2}-2\tilde
  y_{1}\frac{L^{3}}{\tilde b^{2}}< b^{2}
\ ,
\end{equation}
since
\begin{displaymath}
 \tilde c^{2}=\tilde b^{2}-2\tilde
  y_{1}L+L^{2}<L^{2}\Leftrightarrow L^{2}-2\tilde
y_{1}\frac{L^{3}}{\tilde b^{2}} <0 
\end{displaymath}
and
\begin{equation}
  \label{b.9}
  b^{2}=L^{2}\frac{L^{2}}{\tilde b^{2}}>L^{2}\ .
\end{equation}
Since $b>L$ and $b>c$, $Y\in \mathcal{B}$. Finally, let us look at how
the whole integral transforms. First inserting the transformed
variables including the Jacobian (\ref{b.5}) and second using the homogeneity (\ref{b.2}), we arrive at
\begin{equation}
  \label{b.10}
  \int\limits_{Y\in\mathcal{B}}f(a=L,b,c) = \int\limits_{\tilde Y\in
  \mathcal{A}}\left(\frac{L}{\tilde b}\right)^{2D}f\left(L,\frac{L^{2}}{\tilde
  b},\frac{\tilde c L} {\tilde b}\right)=
 \int\limits_{\tilde Y\in
  \mathcal{A}}\left(\frac{L}{\tilde b}\right)^{2D-\lambda}f(\tilde b,\tilde
  a=L,\tilde c)\ .
\end{equation}
Let us restate the above calculation in a completely geometric interpretation as
sketched in the figure. The mapping consists of two steps:
\begin{itemize}
\item The rescaling with respect to $O$ by a factor $L/b$ which maps $Y$ onto
  $Y^{\prime}$ and $H$ onto $H^{\prime}$. (\ref{b.2}) implies that $f$ is
  changed by a factor $\left(\frac{L}{b}\right)^{-\lambda}$.
\item A mirror operation, which maps $Y^{\prime}$ onto $H$ and $H^{\prime}$
  onto $\tilde Y$, leaving invariant the origin $O$. This operation is a
  permutation of the first two arguments of $f$.
\end{itemize}
The mapping $Y \to \tilde Y$ corresponds to the special conformal
transformation (\ref{b.3}).\\
Analogously, we find that
\begin{equation}
  \label{b.11}
  \int\limits_{Y\in\mathcal{C}}f(a=L,b,c)\ =\ \int\limits_{\tilde Y\in
  \mathcal{A}}\left(\frac{L}{\tilde c}\right)^{2D-\lambda}f(\tilde c,\tilde
  b,\tilde a=L)\ .
\end{equation}


\section*{Acknowledgements}
It is a pleasure to thank  R.\ Blossey, F.\ David, H.W.\ Diehl, 
M.\ Kardar,  and L.\ Sch\"afer  for useful discussions. The clarity of
this presentation has  profited from a critical reading by H.W.\ %
Diehl. We are grateful to Andreas Ludwig for  persisting questions,
and his never tiring efforts to understand the limit of $D\to 2$. 
  This work has  been supported  
by the DFG through the Leibniz program Di 378/2-1, under Heisenberg
grant Wi 1932/1-1, and  NSF grant  PHY99-07949.

\setcounter{section}{17} 
\section{References}
\def\refname{} \vspace*{-1.2cm}

\begin{thebibliography}{10}

\bibitem{Schaefer}
L.~Sch\"afer,
\newblock {\em Excluded Volume Effects in Polymer Solutions},
\newblock Springer Verlag, Berlin, Heidelberg, 1999.

\bibitem{BlatterFeigelmanGeshkenbeinLarkinVinokur1994}
G.~Blatter, M.V. {Feigel'man}, V.B. Geshkenbein, A.I. Larkin  and V.M. Vinokur,
\newblock {\em Vortices in high-temperature superconductors},
\newblock Rev. Mod. Phys. {\bf 66} (1994)   1125.

\bibitem{KPZ}
M.~Kardar, G.~Parisi  and Y.-C. Zhang,
\newblock {\em Dynamic scaling of growing interfaces},
\newblock Phys. Rev. Lett. {\bf 56} (1986)   889--892.

\bibitem{WieseHabil}
K.J. Wiese,
\newblock {\em Polymerized membranes, a review}.
\newblock {\em {\em Volume}~19} of {\em Phase Transitions and Critical
  Phenomena}, Acadamic Press, London, 1999.

\bibitem{Abraham1980}
D.B. Abraham,
\newblock {\em Solvable model with a roughening transition for a planar ising
  ferromagnet},
\newblock Phys. Rev. Lett. {\bf 44} (1980)   1165--1168.

\bibitem{Upton1999}
P.J. Upton,
\newblock {\em Exact interface model for wetting in the planar ising model},
\newblock Phys. Rev. E {\bf 60} (1999)   3475--3478.

\bibitem{NakanishiFisher1982}
H.~Nakanishi and M.E. Fisher,
\newblock {\em Multicriticality of wetting, prewetting, and surface
  transition},
\newblock Phys. Rev. Lett. {\bf 49} (1982)   1565--1568.

\bibitem{BrezinHalperinLeibler1983}
B.I.~Halperin E.~Br\'ezin and S.~Leibler,
\newblock {\em Critical wetting in three dimensions},
\newblock Phys.~Rev. Lett. {\bf 50} (1983)   1387.

\bibitem{LipowskyKrollZia1983}
R.~Lipowsky, D.M. Kroll  and R.K.P. Zia,
\newblock {\em Effective field theory for interface delocalization
  transitions},
\newblock Phys. Rev. B {\bf 27} (1983)   4499--4502.

\bibitem{KrollLipowskyZia1985}
D.M. Kroll, R.~Lipowsky  and R.K.P. Zia,
\newblock {\em Universality classes for critical wetting},
\newblock Phys. Rev. B {\bf 32} (1985)   1862.

\bibitem{FisherHuse1985}
D.S. Fisher and D.A. Huse,
\newblock {\em Wetting transitions: a functional renormalization-group
  approach},
\newblock Phys. Rev. B {\bf 32} (1985)   247--56.

\bibitem{LipowskyFisher1987}
R.~Lipowsky and M.E. Fisher,
\newblock {\em Scaling regimes and functional renormalization for wetting
  transitions},
\newblock Phys. Rev. B {\bf 36} (1987)   2126--2241.

\bibitem{DavidLeibler1990}
F.~David and S.~Leibler,
\newblock {\em Multicritical unbinding phenomena and nonlinear functional
  renormalization group},
\newblock Phys. Rev. B {\bf 41} (1990)   12926--9.

\bibitem{ForgasLipowskyNieuwenhuizenInDombGreen}
G.~Forgas, R.~Lipowsky  and T.M. Nieuwenhuizen,
\newblock {\em The behaviour of interfaces in ordered and disordered systems}.
\newblock {\em {\em Volume}~14} of {\em Phase Transitions and Critical
  Phenomena}, pages 136--376, Academic Press London, 1991.

\bibitem{Duplantier1989}
B.~Duplantier,
\newblock {\em Interaction of crumpled manifolds with euclidean elements},
\newblock Phys. Rev. Lett. {\bf 62} (1989)   2337.

\bibitem{DDG1}
F.~David, B.~Duplantier  and E.~Guitter,
\newblock {\em Renormalization of crumpled manifolds},
\newblock Phys. Rev. Lett. {\bf 70} (1993)   2205.

\bibitem{DDG2}
F.~David, B.~Duplantier  and E.~Guitter,
\newblock {\em Renormalization theory for interacting crumpled manifolds},
\newblock Nucl. Phys. {\bf B 394} (1993)   555--664.

\bibitem{KantorNelson1987a}
Y.~Kantor and D.R. Nelson,
\newblock {\em Crumpling transition in polymerized membranes},
\newblock Phys. Rev. Lett. {\bf 58} (1987)   2774--2777.

\bibitem{KantorNelson1987b}
Y.~Kantor and D.R. Nelson,
\newblock {\em Phase transitions in flexible polymeric surfaces},
\newblock Phys. Rev. {\bf A 36} (1987)   4020--4032.

\bibitem{KantorKardarNelson1986a}
Y.~Kantor, M.~Kardar  and D.R. Nelson,
\newblock {\em Statistical mechanics of tethered surfaces},
\newblock Phys. Rev. Lett. {\bf 57} (1986)   791--795.

\bibitem{KantorKardarNelson1986b}
Y.~Kantor, M.~Kardar  and D.R. Nelson,
\newblock {\em Tethered surfaces: Statics and dynamics},
\newblock Phys. Rev. {\bf A 35} (1987)   3056--3071.

\bibitem{KardarNelson1987}
M.~Kardar and D.R. Nelson,
\newblock {\em $\varepsilon$ expansions for crumpled manifolds},
\newblock Phys. Rev. Lett. {\bf 58} (1987)   1289 and 2280 E.

\bibitem{AronovitzLubensky1988}
J.A. Aronovitz and T.C. Lubensky,
\newblock {\em Fluctuations of solid membranes},
\newblock Phys. Rev. Lett. {\bf 60} (1988)   2634--2637.

\bibitem{DDG3}
F.~David, B.~Duplantier  and E.~Guitter,
\newblock {\em Renormalization and hyperscaling for self-avoiding manifold
  models},
\newblock Phys. Rev. Lett. {\bf 72} (1994)   311.

\bibitem{DDG4}
F.~David, B.~Duplantier  and E.~Guitter,
\newblock {\em Renormalization theory for the self-avoiding polymerized
  membranes},
\newblock cond-mat\slash {\bf 9702136} (1997).

\bibitem{DavidWiese1996}
F.~David and K.J. Wiese,
\newblock {\em Scaling of self-avoiding tethered membranes: 2-loop
  renormalization group results},
\newblock Phys. Rev. Lett. {\bf 76} (1996)   4564.

\bibitem{WieseDavid1997}
K.J. Wiese and F.~David,
\newblock {\em New renormalization group results for scaling of self-avoiding
  tethered membranes},
\newblock Nucl. Phys. {\bf B 487} (1997)   529--632.

\bibitem{BowickTravesset2001}
G.~Thorleifsson M.~Bowick, A.~Cacciuto and A.~Travesset,
\newblock {\em Universality classes of self-avoiding fixed-connectivity
  membranes},
\newblock Eur. Phys. J. {\bf E 5} (2001)   149.

\bibitem{PinnowWieseProgress}
H.~Pinnow and K.J. Wiese,
\newblock work in progress.

\bibitem{WieseDavid1995}
K.J. Wiese and F.~David,
\newblock {\em Self-avoiding tethered membranes at the tricritical point},
\newblock Nucl. Phys. {\bf B 450} (1995)   495--557.

\bibitem{Hwa1990}
T.~Hwa,
\newblock {\em Generalized $\varepsilon$ expansion for self-avoiding tethered
  manifolds},
\newblock Phys. Rev. {\bf A 41} (1990)   1751--1756.

\bibitem{LassigLipowsky1993}
M.~L\"assig and R.~Lipowsky,
\newblock {\em Critical roughening of interfaces: a new class of renormalizable
  field theories},
\newblock Phys. Rev. Lett. {\bf 70} (1993)   1131--4.

\bibitem{Goulian1991}
M.~Goulian,
\newblock {\em The gaussian approximation for self-avoiding tethered surfaces},
\newblock J. Phys. II France {\bf 1} (1991)   1327--1330.

\bibitem{LeDoussal1992}
P.~Le Doussal,
\newblock {\em Tethered membranes with long-range self-avoidance: large
  dimension limit},
\newblock J. Phys. {\bf A 25} (1992)   469--476.

\bibitem{GuitterPalmeri1992}
E.~Guitter and J.~Palmeri,
\newblock {\em Tethered membranes with long-range interaction},
\newblock Phys. Rev. {\bf A 45} (1992)   734--744.

\bibitem{DavidWiese1998}
F.~David and K.J. Wiese,
\newblock {\em Large orders for self-avoiding membranes},
\newblock Nucl. Phys. {\bf B 535} (1998)   555--595.

\bibitem{DiehlInDombGreen}
H.W. Diehl,
\newblock {\em Field-theoretical approach to critical behaviour of surfaces}.
\newblock {\em {\em Volume}~10} of {\em Phase Transitions and Critical
  Phenomena}, pages 76--267, Academic Press London, 1986.

\bibitem{EisenrieglerKremerBinder1982}
E.~Eisenriegler, K.~Kremer  and K.~Binder,
\newblock {\em Adsorption of polymer chains at surfaces: scaling and
  Monte-Carlo analysis},
\newblock J. of Chem. Phys. {\bf 77} (1982)   6296.

\end{thebibliography}

\end{appendix}

\end{document}